\documentclass[10pt,aps,prb,reprint,superscriptaddress,floatfix,longbibliography]{revtex4-2}

\usepackage{physics}
\usepackage{amsmath}
\usepackage{amssymb}
\usepackage{amsthm}
\usepackage[dvipsnames]{xcolor}

\usepackage{amsbsy}
\usepackage{amstext}
\usepackage{graphicx}

\usepackage{color}
\usepackage{mathtools}
\usepackage{subfigure}
\usepackage{multirow}

\usepackage{amsfonts}
\usepackage{dcolumn}
\usepackage{bbold,bm}
\usepackage[bookmarks=false,linkcolor=Orange,urlcolor=MidnightBlue,colorlinks,citecolor=Maroon]{hyperref}

\makeatletter
\newcommand{\D}{\Delta}

\newcommand{\gsf}{\Gamma_{\text{sf}}}
\newcommand{\gin}{\Gamma_{\text{in}}}

\makeatother
\usepackage{babel}

\begin{abstract}
We consider a planar superconducting-normal-metal (SN) junction with both inelastic and spin-flip scattering processes present. In the diffusive limit, we use a one-dimensional formulation of the Usadel equation to compute the self-consistent energy dependence of the single-particle density of states as a function of distance from the interface on both the superconducting and metallic sides for various spatial profiles of a pair-breaking spin-flip term. The pair-breaking processes fill in the superconducting gap at zero energy, which is reflected in the zero-bias tunneling conductance in scanning tunneling microscopy/spectroscopy experiments, in the vicinity of the junction. We also investigate the impact of having a partially transparent interface at the junction. We compare our findings with the observed exponential rise in the zero-bias conductance at the 1H step edge in recent experiments on 4Hb-TaS$_2$ [A. K. Nayak {\em et al.}, Nat. Phys. {\bf 17}, 1413 (2021)].
\end{abstract}

\begin{document}
\allowdisplaybreaks

\title{Self-consistent evaluation of proximity and inverse proximity effects with pair-breaking in diffusive SN junctions}
%\title{Proximity effects in diffusive SN junctions with spatially varying spin-flip scattering}

\author{Arpit Raj}
\email{raj.a@northeastern.edu}
\affiliation{Department of Physics, Northeastern University, Boston, Massachusetts 02115, USA}

\author{Patrick A. Lee}
\affiliation{Department of Physics, Massachusetts Institute of Technology, Cambridge, MA 02139, USA}

\author{Gregory A. Fiete}
\affiliation{Department of Physics, Northeastern University, Boston, Massachusetts 02115, USA}
\affiliation{Department of Physics, Massachusetts Institute of Technology, Cambridge, MA 02139, USA}

\maketitle

% --------------------------------------------------------------------------
\section{Introduction}

When a superconductor (S) and a normal metal (N) are in close electrical contact, the metal can acquire superconducting correlations which can extend into the N side over a characteristic length scale, typically on the order of the superconducting coherence length. This is known as the proximity effect~\cite{deGennes:rmp64}. Conversely, the superconductivity on the S side of the system can be influenced by the presence of the normal metal, a phenomenon known as the inverse proximity effect, which can lead to a reduction in the superconducting gap close to the interface. Both effects can be important in hybrid superconductor-normal metal (SN) nanostructures~\cite{pannetier2000}. The proximity and inverse proximity effects are influenced by factors such as the strength of the superconducting pairing, symmetry of the order parameter, spatial dimensionality of the two sides, and the quality of the SN interface. Understanding these effects is crucial for the design and optimization of superconducting devices and materials. Proximity effects are usually observed through a spatially varying local density of states (LDOS) across the interface and have been probed using local tunneling measurements, possible with a scanning tunneling microscope (STM), in the past~\cite{Gueron:prl96,Moussy:EPL2001,vinet2001,meschke2011}.

On the theoretical side, proximity and inverse proximity effects in SN junctions are commonly studied in the diffusive limit using a quasiclassical theory governed by the Usadel equations~\cite{Usadel:prl70}. This approach is particularly useful when the spatial dependence of the gap parameter and local density of states (LDOS) are of interest. Another advantage of this formalism is that it allows for a straightforward inclusion of both inelastic and spin-flip (pair-breaking) scattering processes by using the corresponding scattering rates, $\gin$ and $\gsf$. While the inelastic scattering provides a homogeneous energy broadening throughout the system (say, from thermal fluctuations in a system at uniform temperature), the spin-flip scattering can be highly position dependent. This may simply result, for instance, from presence of different magnetic impurities on the S and N sides. Surprisingly, although several authors have studied SN junctions using the Usadel equations in the past~\cite{Belzig:prb96,Crouzy:prb05,golubov1997,yip1995,Gueron:prl96}, the effect of spatial profiles for $\gsf$ has not been studied to the best of our knowledge.

We aim to bridge this gap by presenting a detailed study of the dependence of the superconducting gap $\D$, zero-bias conductance (ZBC) and tunneling spectra across a diffusive SN junction on the strength and spatial profile of the spin-flip scattering for a range of parameters characterizing the materials and interface transparency.  We employ a fully self-consistent solution of the Usadel equations to determine the proximity and inverse proximity effects. We also consider various limits that can be treated analytically. 

A second impetus of this work is to revisit the differential tunneling conductance, $dI/dV$, spectra seen in recent STM experiments on the candidate topological superconductor 4Hb-TaS$_2$~\cite{nayak2021} which is believed to have spin-flip scattering centers and SN junctions on surface step edges.  The model we consider assumes a trivial superconducting state and thus it allows us to explore whether any of the experimental features can be captured without assuming the presence of topological boundary modes.  Because unambiguous identification of topological superconductors has proven elusive, it appears that intensive theoretical modeling combined with multiple experimental probes pointing to the same topological state will be needed to reach a definitive conclusion on any given material \cite{raj2023nonlinear}.

In this context, the transition metal dichalcogenide 4Hb-TaS$_2$ has emerged as a promising candidate for unconventional superconductivity with several experimental studies suggesting the existence of spontaneous time-reversal breaking and/or topological superconductivity~\cite{ribak2020, nayak2021, persky2022, almoalem2022, silber2024,raj2023nonlinear}. Its unit cell is comprised of an alternate stacking of 1H-TaS$_2$ and 1T-TaS$_2$ layers, and the transition temperature is  $T_c=2.7$ K~\cite{ribak2020, nayak2021, silber2024}. It is worth noting that while bulk 2H-TaS$_2$ is a superconductor with a critical temperature of $0.7$ K~\cite{nagata1992}, $T_c$ is progressively enhanced for thinner flakes, reaching $2.2$ K for 5 layers~\cite{navarro2016enhanced}.

Part of the motivation for the expectation of unconventional superconductivity is that  1T-TaS$_2$ was proposed to host a quantum spin liquid state~\cite{law2017,he2018} in bulk crystals. This idea has received support from   STM data on monolayers of the closely related compounds 1T-TaSe$_2$~\cite{ruan2021evidence} and 1T-NbSe$_2$~\cite{houquantum2024}.   Theoretical studies suggest that the ground state of a Hubbard model near the Mott transition is  a chiral spin liquid phase~\cite{hu2016,szasz2020}. On the other hand, band calculations~\cite{crippa2024,wen2021}  point to the importance of charge transfer from the 1T to 1H layer in 4Hb bulk crystals, leading to the depletion of local moments and resulting in a  metallic state in the 1T layer. There is direct evidence from STM tunneling that this is the case, at least for the top layer~\cite{nayak2023}. If this continues to be the case in the bulk, the motivation for unconventional pairing in the 1H layers driven by a spin liquid state in the 1T layers is not so clear.

Furthermore, the 4Hb superconductors are known to be in the dirty limit~\cite{fischer2023}, and most known unconventional  superconductors usually do not survive in this limit. We note that a recent paper~\cite{fischer2023} has put forward a conventional explanation for the $\pi$ phase shift in a Little-Parks experiment for 4Hb-TaS$_2$ rings~\cite{almoalem2022}. We are therefore motivated to learn if an alternative explanation of the edge tunneling data~\cite{nayak2021} is possible.

\begin{figure}[t!]
    \centering
    \includegraphics[scale=0.47]{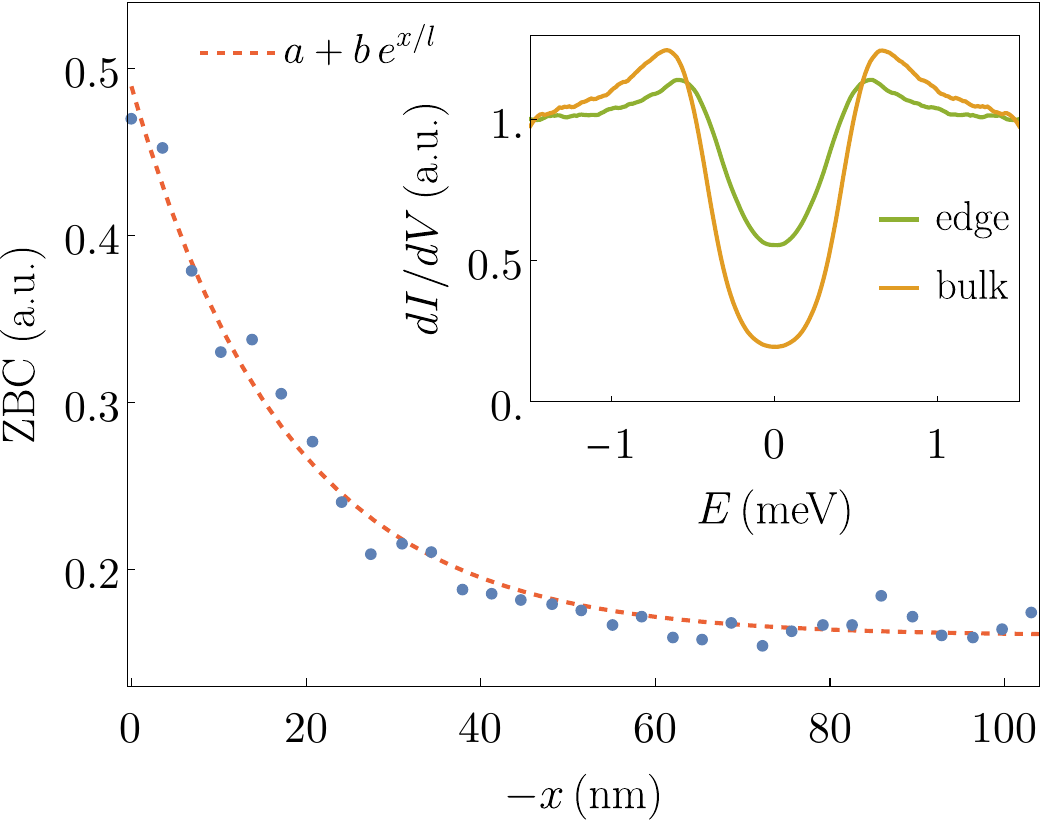}
    \caption{Experimental results for the 1H step edge reproduced from Ref.~\cite{nayak2021}. Main panel shows the zero bias peak decaying from the interface at $x=0$ into the superconducting 1H side ($x<0$). Dashed red line is a fit to an exponential form with fit parameters: $a=0.16$, $b=0.329$, $l=17.8$ nm. Inset shows the tunneling spectra at the edge and in the bulk.} 
    \label{fig:exp}
\end{figure}
\begin{figure}[t!]
    \centering
    \includegraphics[scale=0.47]{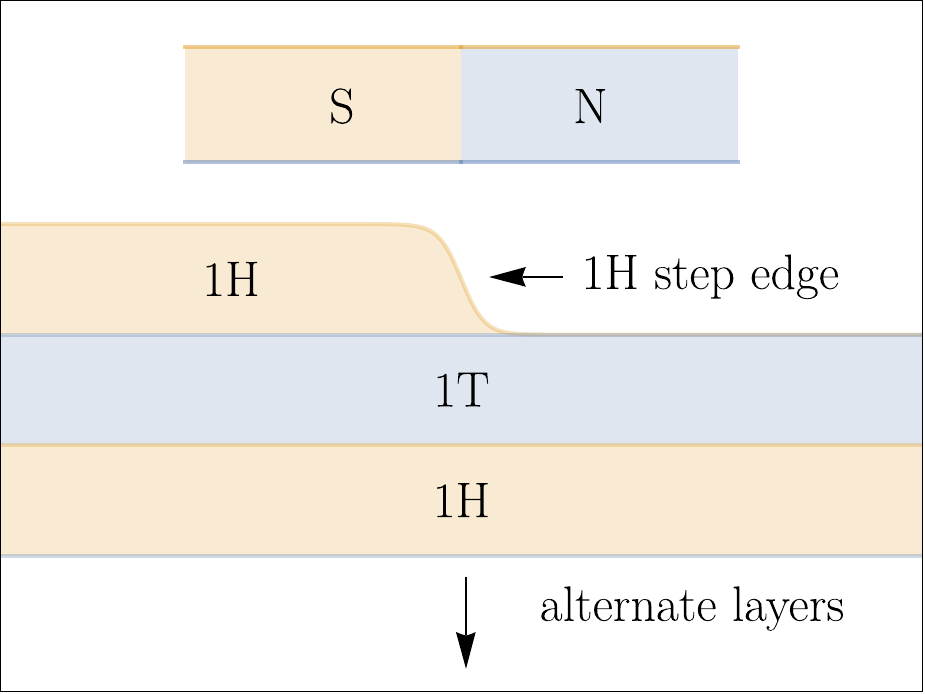}
    \caption{Schematic of the 1H step edge configuration used in the experiment~\cite{nayak2021}. Top drawing shows the simplified SN junction geometry we used to model it. Similar experimental results are seen where the role of 1T and 1H are reversed--switching them produces a 1T step edge.}
    \label{fig:step}
\end{figure}

In this work, we treat the 1T layer in the 4Hb compound as a normal metal (assuming screening/charge transfer from the adjacent 1H layer~\cite{nayak2023,crippa2024,wen2021}) and focus on the experimental results of Ref.~\cite{nayak2021} where a spatially uniform residual DOS is observed within the superconducting gap throughout the 1H surface termination and an exponential rise in the ZBC was seen at the 1H step edge. We have reproduced these results in Fig.~\ref{fig:exp}. The 1H step edge geometry is shown in Fig.~\ref{fig:step}. To explain these observations, we model the 1H step edge as an SN junction and calculate the $\dd{I}/\dd{V}$ spectra across it by solving the Usadel equations self-consistently. The model should apply to the experimental configuration shown in Fig.~\ref{fig:step} as long as the tunneling between 1T and 1H layers across the step edge dominate over the tunneling between layers in the bulk.

We note that experimentally a similar ZBC was seen in a second geometry where the 1H later is at the bottom of a crater surrounded by a 1T layer (step edge). We believe our model will also produce similar results for this geometry because  quasi-particles will extend from 1T into the 1H layer in the same way, the only difference being that the states under the 1H layer are not accessible to STM tunneling.  By varying the phenomenological parameters of the Usadel equations were are able to match many features of the data. We find that a rather broad choice of parameters give similar behavior for the ZBC. Further theoretical studies and additional STM data particularly on the 1T side may be needed to test the applicability of our model.
  
Our paper is organized as follows. Section~\ref{sec:2} describes the model and the Usadel theory along with features of its non-self-consistent solutions. In Section~\ref{sec:3}, we present and discuss the results of fully self-consistent calculations, focusing on the zero-bias conductance (ZBC) and local density of states (LDOS). In Section~\ref{sec:4}, we present results that reproduce some features of the data measured in 4Hb-TaS$_2$. We finish with our main conclusions in Section~\ref{sec:conclusion}.

% --------------------------------------------------------------------------
\section{Model and Usadel equations}
\label{sec:2}
% --------------------------------------------------------------------------

We consider an SN junction with the superconductor on the $x<0$ side and the normal metal on the $x>0$ side. The system is taken to be semi-infinite on both the superconducting and normal sides to avoid ``mini-gap" features associated with quantum confinement effects from a finite length system on either side of the junction. The superconductor and normal metal have diffusion constants $D_S$ and $D_N$, respectively. Without any pair breaking terms, the superconductor has a zero-temperature gap of $\D_0$ and a transition temperature $T_{cS}$ (we will simply call it $T_c$). The normal metal is non-superconducting ($T_{cN}=0$). Unless stated otherwise, the system temperature is taken to be zero. The proximity lengths, $\xi_{S/N} = \sqrt{\hbar D_{S/N}/2\D_0}$, define the length scales on the S and N sides.

To study the spatial dependence of $\D$ and the LDOS across this junction, we use the one-dimensional Usadel equation with inelastic and spin-flip scattering terms, given by~\cite{Usadel:prl70, Belzig:prb96, Crouzy:prb05, Hammer:prb07},
\begin{align}
\begin{split}
&\frac{\hbar D}{2} \pdv[2]{\theta}{x} + \left[iE-\gin-2\gsf(x)\cosh{\theta}\right]\sinh{\theta}\\
&\hspace{4.5cm} - i\D(x)\cosh{\theta} = 0,
\end{split}\label{eq:Usadel_full}
\end{align}
where $\theta(x,E)$ is a function of the distance to the interface, $x$, and the energy, $E$. Here $D$ is the diffusion constant, $\gin$ is an energy broadening associated with inelastic scattering, and $\gsf$ is an energy broadening associated with spin flip scattering. While $\gin$ will be taken to be constant (spatially uniform) throughout the system, various profiles for $\gsf(x)$ will be considered. The position dependent order parameter, $\D(x)$, is obtained from the self-consistency condition~\cite{Moussy:EPL2001,Cherkez:prx14}:
\begin{equation}
\D(x) = NV \int_0^{E_c} \dd{E} \,\tanh(\frac{E}{2k_BT}) \text{Re}\left\{\sinh{\theta}\right\}. \label{eq:gap_full}
\end{equation}
Here $E_c$ is a cutoff energy ($\sim$ Debye energy), $N$ is the normal-state DOS at the Fermi energy ($N_S$ on the S and $N_N$ on the N side) and $V$ is the magnitude of the effective electron-phonon coupling (calculated at zero temperature and without any pair breaking processes) in units such that $NV$ is dimensionless. While $V_S$ is finite, $V_N=0$ which means $\D(x>0)=0$. Note that we have chosen the order parameter to be real which is possible for the chosen geometry of the SN junction. From the solution of the Usadel equation, the LDOS, $\rho(x,E)$, is obtained as~\cite{Belzig:prb96,Crouzy:prb05}
\begin{equation}
    \rho(x,E) = N\,{\rm Re}\left\{\cosh(\theta(x,E))\right\},
    \label{eq:LDOS}
\end{equation}
and the ZBC is simply $\rho(x,0)$.

Solving Eq.~\eqref{eq:Usadel_full} requires matching the bulk values of $\theta$ on the two ends and satisfying additional constraints (for $\theta$ and $\partial_x\theta$) at the interface (see Appendix~\ref{appendix:selfc}). For an interface of arbitrary transparency, these constraints are~\cite{kuprianov1988}:
\begin{align}
\begin{split}
    \sigma_S\eval{\pdv{\theta(x,E)}{x}}_{x=0^-} &= \sigma_N\eval{\pdv{\theta(x,E)}{x}}_{x=0^+} \\
    &= \frac{\sinh(\theta(0^+,E)-\theta(0^-,E))}{R_{\rm int}\,A_{\rm int}},
\end{split}\label{eq:bc_full}
\end{align}
where $\sigma_S$ and $\sigma_N$ are the normal state conductivity of the superconductor and the normal metal, respectively, and $R_{\rm int}$ and $A_{\rm int}$ are the interface resistance and area, respectively. An important dimensionless parameter, $\gamma$, which combines $\sigma_{S/N}$ and $D_{S/N}$ is often used to characterize the two systems, and is given by 
\begin{equation}
\gamma = \frac{\sigma_N\xi_S}{\sigma_S\xi_N} = \frac{\sigma_N}{\sigma_S} \sqrt{\frac{D_S}{D_N}}.
\end{equation}
This is known as the mismatch parameter. Another dimensionless parameter
\begin{equation}
    \gamma_B = \frac{\sigma_N}{\xi_N}R_{\rm int}\,A_{\rm int} = \gamma\frac{\sigma_S}{\xi_S}R_{\rm int}\,A_{\rm int}
\end{equation}
is used to characterize the interface transparency. When $\gamma_B = 0$, the interface is perfectly transparent whereas for $\gamma_B \rightarrow \infty$, the S and N sides get decoupled and there are no proximity or inverse-proximity effects. Note that, unless stated otherwise, hereafter $x$ will be in units of $\xi_S$ and $\xi_N$ on the S and N sides, respectively.  

A self-consistent solution to the Usadel equation is obtained numerically by an iterative procedure. Features of its solution for a uniform superconductor that are relevant to our study are discussed in Appendix~\ref{appendix:selfc_uniform}. For the non-uniform case of an SN junction, details of the numerical method used to solve it are provided in Appendix~\ref{appendix:selfc}. While self-consistent solutions are necessary when one wishes to match theory against experimental data, it is very useful to have a non-self-consistent analytical solution to gain intuition about how various parameters in the model affect the solution. For $\gin\neq 0$ and $\gsf=0$, an analytical solution is provided in Ref~\cite{Belzig:prb96}. When $\gsf\neq 0$, such a solution is difficult to obtain for an arbitrary spatial profile of the spin-flip scattering. However, by choosing $\gsf(x)=\gsf\Theta(x)$ where $\Theta(x)$ is the Heaviside step function, the Usadel equations on the S and N sides can be linearized and solved analytically for $\gamma_B=0$ and semi-analytically when $\gamma_B$ is finite (details given in Appendix~\ref{appendix:nonselfc}). Note that the technique used here to linearize Eq.~\eqref{eq:Usadel_full} on the S side can easily be extended to SS$'$ junctions having only inelastic scattering processes. 

For $\gamma_B=0$, the ZBC on the S and N sides, obtained from the linearized Usadel equations, is given by  
\begin{align}
\begin{split}
    & \frac{\rho(x<0,0)}{N_S} = \cos \left[\tan ^{-1}\left(\tfrac{\D_0 }{\gin}\right)\left(1-\tfrac{e^{x \sqrt{\sqrt{1+\gin^2/\D_0^2}}}}{1+\frac{1}{\gamma}\sqrt{\frac{\sqrt{\gin^2+\D_0^2}}{\gin+2\gsf}}}\right)\right],
\end{split} \label{eq:zbcS}\\
\begin{split}
    & \frac{\rho(x>0,0)}{N_N} = \cos \left[\tfrac{\tan ^{-1}\left(\frac{\D_0 }{\gin}\right) e^{-x \sqrt{\frac{\gin+2\gsf}{\D_0}}}}{1+\gamma\sqrt{\frac{\gin+2\gsf}{\sqrt{\gin^2+\D_0^2}}}}\right],
\end{split} \label{eq:zbcN}
\end{align}
which have bulk values of $N_S/\sqrt{1+(\D_0/\gin)^2}$ and $N_N$, respectively. We should point out that, based on the self-consistent solution for a uniform superconductor given in Appendix~\ref{appendix:nonselfc}, increasing $\gin$ leads to increase in the bulk value of ZBC on the S side until it hits $N_S$ at $\gin\approx\D_0/2$, beyond which superconductivity is completely destroyed and $\rho(x<0,0)=N_S$ (see Fig.~\ref{fig:uniform_1}(c)). As seen from Eq.~\eqref{eq:zbcS}, the ZBC rises exponentially as we move from the bulk of S towards the interface, with
\begin{align}
    \frac{\rho(0^-,0)}{N_S} = \frac{\rho(0^+,0)}{N_N} = \cos \left[\frac{\tan ^{-1}\left(\frac{\D_0 }{\gin}\right)}{1+\gamma\sqrt{\frac{\gin+2\gsf}{\sqrt{\gin^2+\D_0^2}}}}\right].
\end{align} 
This clearly shows that increasing $\gamma$ and/or $\gsf$ leads to a rise in the ZBC. 

For $\gamma_B\neq 0$, the bulk values of the ZBC remain the same as expected. However, $\gamma_B$ inversely affects the rise in ZBC on the S side, which is easy to understand. Larger $\gamma_B$ leads to smaller proximity and inverse proximity effects which means that the bulk value of the ZBC on either sides will undergo less change compared to $\gamma_B=0$ case. This can also be seen from the analytical solutions given in Appendix~\ref{appendix:nonselfc}. Based on these solutions, we also note that when $\gamma_B$ is nonzero, $\rho(0^-,0)/N_S\neq \rho(0^+,0)/N_N$. This can be a useful guide in deciding whether one needs to consider $\gamma_B$ when modelling a given experimental data set.

% --------------------------------------------------------------------------
\section{Results of self-consistent calculation}
\label{sec:3}

\begin{figure}[t!]
    \centering
    \includegraphics[scale=0.4]{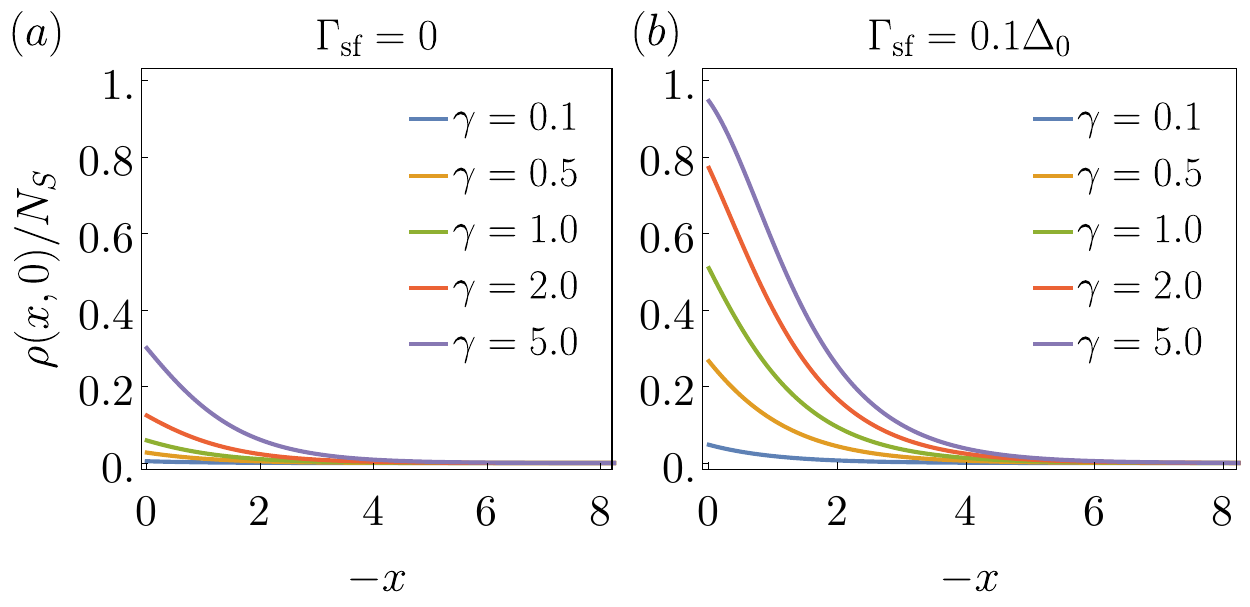}
    \caption{Spatial variation of the ZBC on the S side without (a), and with (b) a step spin-flip scattering term.}
    \label{fig:zbc1}
\end{figure}
\begin{figure}[ht]
    \centering
    \includegraphics[scale=0.4]{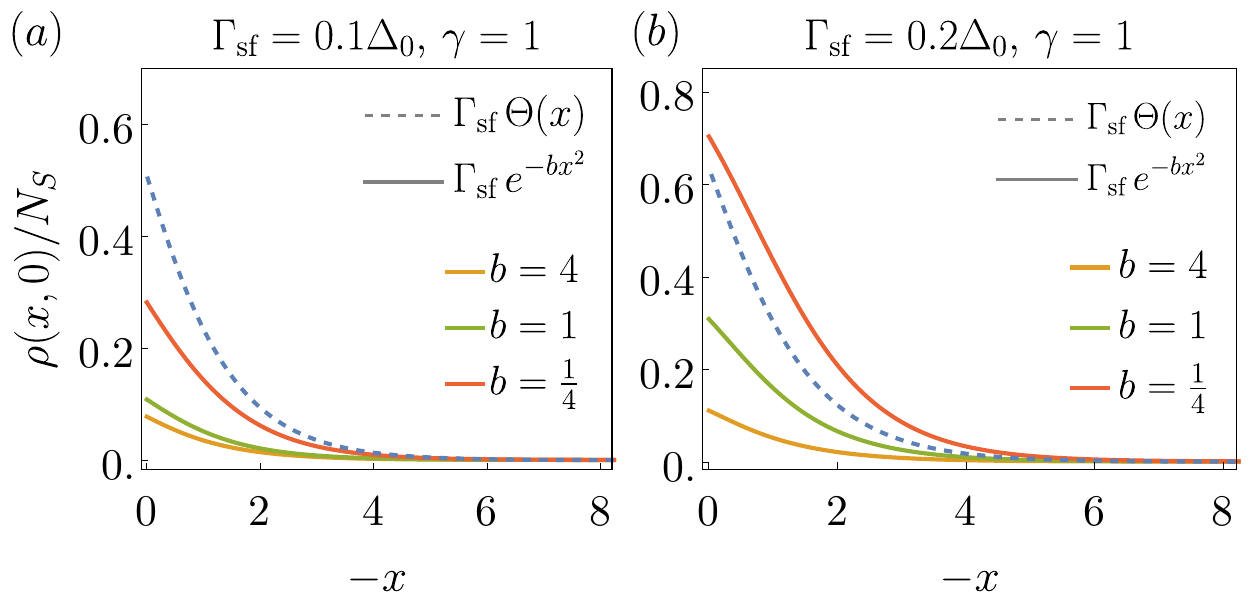}
    \caption{Comparison between the ZBC on the S side for step and Gaussian spin-flip scattering profiles.}
    \label{fig:zbc_compare}
\end{figure}
\begin{figure}[t!]
    \centering
    \includegraphics[scale=0.4]{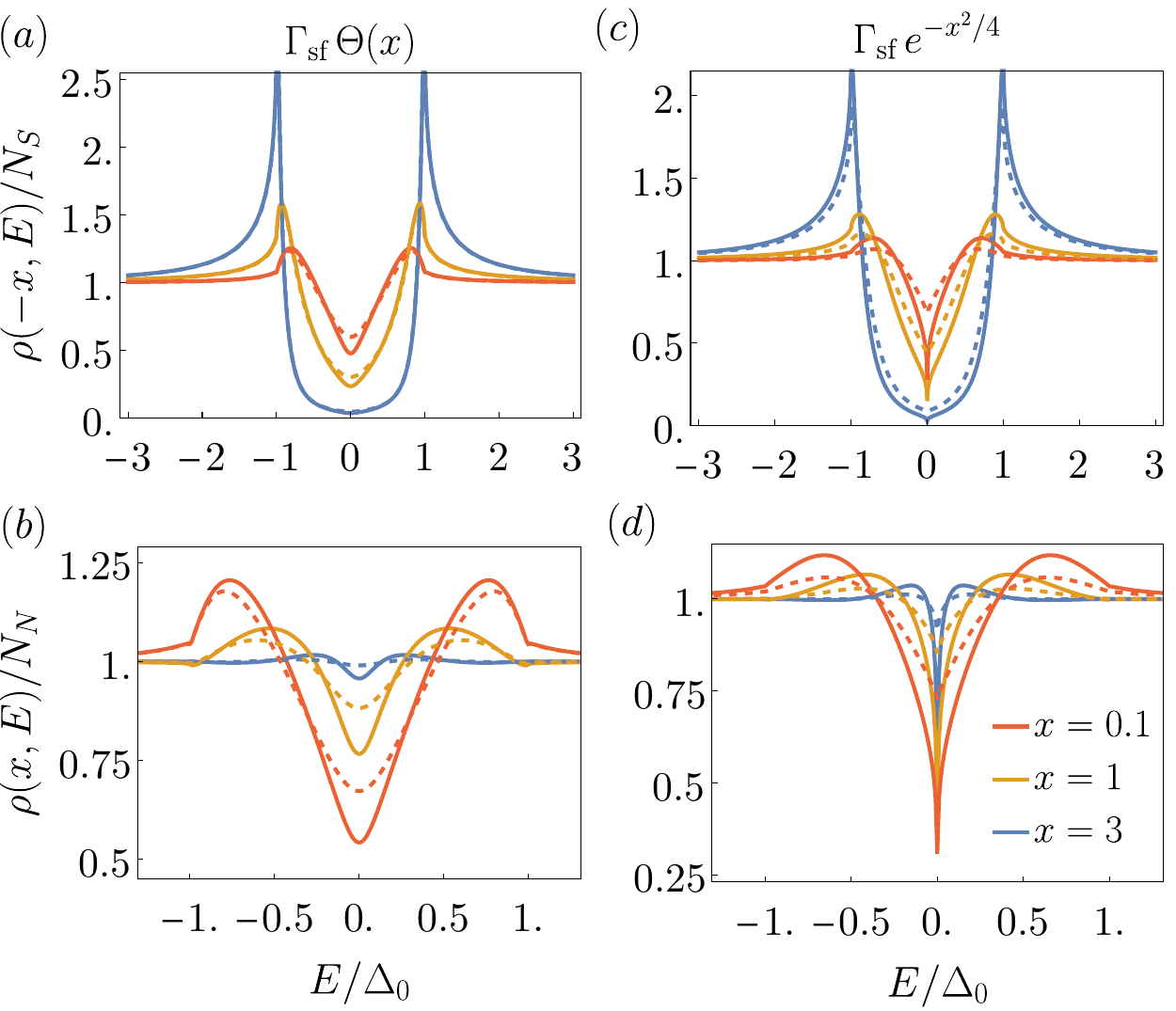}
    \caption{Comparison between LDOS for $\gsf=0.1\D_0$ (solid) and $\gsf=0.2\D_0$ (dashed) on the S side (upper) and N side (lower) for different spatial profiles of the spin-flip scattering term. We have set $\gamma=1$.}
    \label{fig:ldos_compare}
\end{figure}
Having gained insights from the non-self-consistent solutions in the previous section, we now present results of self-consistent calculations. Note that, for numerical stability of the self-consistent solution of the Usadel equations, we took a tiny value for the inelastic term, $\gin=0.001\D_0$, throughout this section. We first consider the case $\gamma_B=0$. In addition to the step $\gsf\,\Theta(x)$, we also consider a Gaussian spin-flip scattering term $\gsf\,e^{-bx^2}$.

Results for the step profile are shown in Fig.~\ref{fig:zbc1} which follows the expected behavior with increasing $\gsf$ and $\gamma$ ({\em i.e.}, an increase in the ZBC). Next, we compare the ZBC for the step and Gaussian profiles for $\gamma=1$. Results are shown in Fig.~\ref{fig:zbc_compare}. We see a significant difference between the Gaussian and step cases, with the former giving rise to a much smaller rise in ZBC compared to the latter unless one takes very large values of $\gsf$ and small $b$. A reduction in $b$ causes the spin-flip processes to penetrate further into the S side leading to greater suppression of superconductivity (hence larger increase in ZBC). 
The difference between these two profiles becomes even more evident from the LDOS plots, shown in Fig.~\ref{fig:ldos_compare}. Around $E=0$, we see an $E^2$ dependence for $\rho$ in the step case, as predicted from the linearized solution (Appendix~\ref{appendix:nonselfc}). However, the Gaussian spin-flip term gives rise to a cusp in the LDOS around zero energy. This is not unexpected if we note that for $\gin=\gsf=0$, we get an $\sqrt{|E|}$ behavior near $E=0$ (Appendix~\ref{appendix:nonselfc}). Introducing a symmetric or interface-localized spin-flip scattering term should not remove this kink while $\gin=0$ and $\gsf$ is not too large.

\begin{figure}[t]
    \centering
    \includegraphics[scale=0.4]{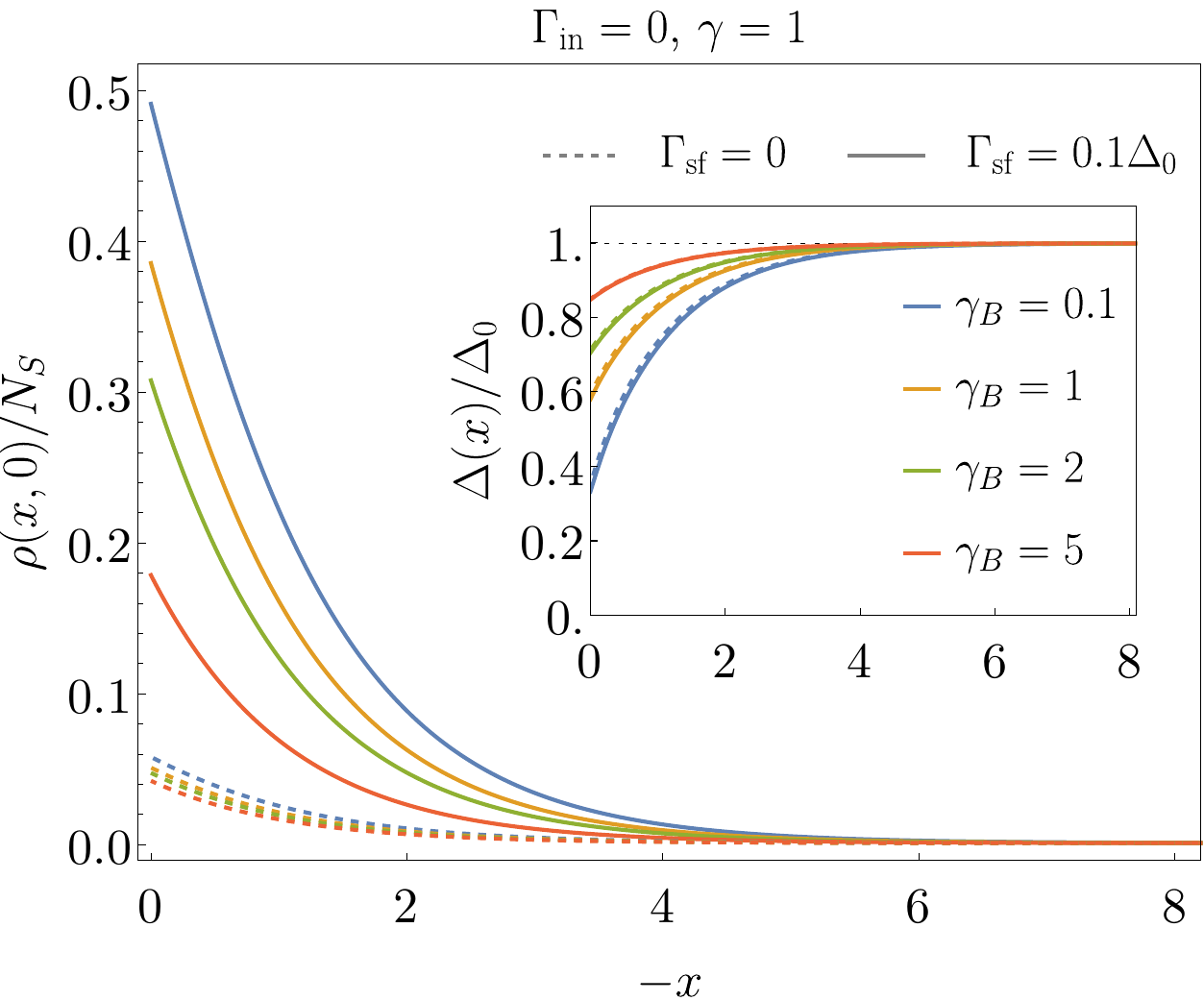}
    \caption{Effect of interface transparency on the ZBC on the S side, with (solid) and without (dashed) a step-like spin-flip scattering term $\gsf\Theta(x)$. The inset shows the corresponding self-consistent gap $\D(x)$.}
    \label{fig:zbcB}
\end{figure}

\begin{figure*}[t]
    \centering
    \includegraphics[scale=0.42]{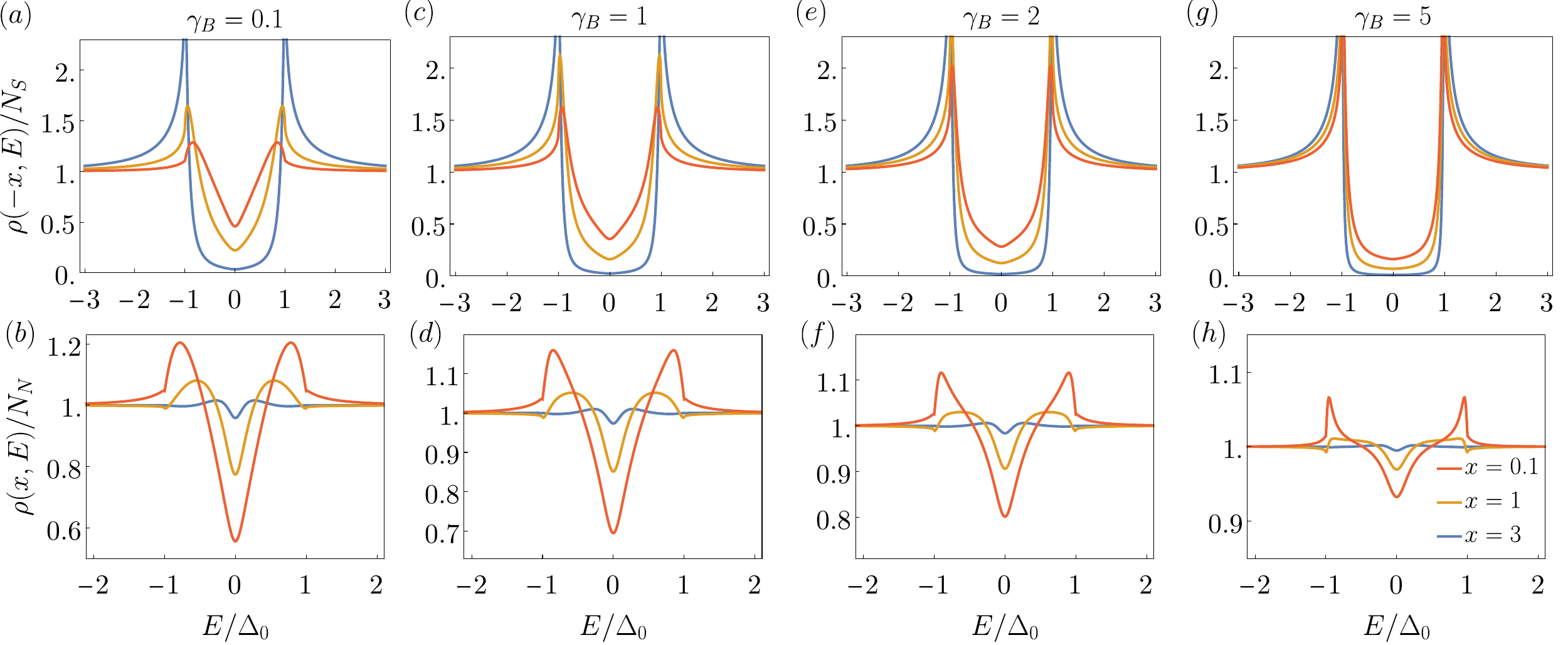}
    \caption{LDOS on the S side (upper) and N side (lower) for different interface transparency. We have used $\gamma=1$, $\gsf=0.1\D_0$.}
    \label{fig:ldos_compare_B}
\end{figure*}

Finally, we look at the effect of having a partially transparent interface ($\gamma_B\neq 0$). We only consider the step profile here as there is no reason to expect other cases will not show similar behavior with finite $\gamma_B$. Results for the ZBC are shown in Fig.~\ref{fig:zbcB}. As expected, we see a decrease in the ZBC as $\gamma_B$ is increased. Also, large $\gamma_B$ leads to reduced proximity and inverse-proximity effects and this is clearly seen from the LDOS plots in Fig.~\ref{fig:ldos_compare_B}. 

% --------------------------------------------------------------------------
\section{Results for $4\text{Hb-TaS}_2$}
\label{sec:4}

\begin{table}[b]
    \caption{Various combinations of $(\gamma_B,\gamma,\gsf)$ that reproduce the experimentally observed rise in ZBC near the 1H step edge at $T=0.38$K. We have used a step-like spin-flip scattering term and fixed $\gin=0.16\D_0$.}
    \begin{ruledtabular}
    \begin{tabular}{|m{0.0\textwidth}m{0.08\textwidth}|m{0.08\textwidth}|m{0.08\textwidth}|}
        &$\gamma_B$ & $\gamma$ & $\gsf/\D_0$ \\
         \hline
        &2 & 1 & 0.3 \\
        &1.3 & 1 & 0 \\
        &0.5 & 0.5 & 0.3 \\
        &0.1 & 0.5 & 0.1\\
        &0 & 0.4 & 0.2 \\
        &0 & 0.3 & 0.5 \\
    \end{tabular}
    \end{ruledtabular}
    \label{table:4HbTaS2}
\end{table}
We model the 1H step edge region in Ref.~\cite{nayak2021} as an SN junction (see Fig.~\ref{fig:step}) and take $T=0.38$K, $\D_0=0.44\,$meV and $E_c=100\D_0$. As shown in the inset of Fig.~\ref{fig:uniform_1}(a) in Appendix~\ref{appendix:selfc_uniform}, one cannot explain the observed ZBC at this temperature while keeping $\gin=\gsf=0$. We take $\gin=0.16\D_0$, which is about $2k_BT$. This choice of $\gin$ produces a ZBC of about $0.2N_S$ as is seen in the fit to the symmetrized, averaged, and normalized $dI/dV$ spectra in the bulk of a 1H step (see Fig.~\ref{fig:exp} or the supplementary of Ref.~\cite{nayak2021} for more details). 

We note that getting this bulk offset in the ZBC by taking a finite $\gsf$ on the 1H side is also possible but not desirable. As shown in Fig.~\ref{fig:uniform_1}(c) in Appendix~\ref{appendix:selfc_uniform}, at low temperatures, the ZBC in the bulk rises sharply in a small region when approaching the critical $\gsf$. This will give a tiny bulk value of $\D$ away from the 1H step-edge which is not seen in the experiment. Keeping this in mind, we will only include a spin-flip scattering term on the 1T side (a step profile). With this setup, we solve the Usadel equation self-consistently for different combinations of the mismatch parameter ($\gamma$), the strength of spin-flip scattering ($\gsf$) and the boundary transparency ($\gamma_B$), and obtain the LDOS using Eq.~\eqref{eq:LDOS}. To compare our results with the experiment, we calculate the differential conductance from $\rho$ using
\begin{align}
    \dv{I}{V}\, (x,\widetilde{V}) &= -\int_{-\infty}^{+\infty} \dd{E}\, \rho(x,E)\pdv{n_F(E-e\widetilde{V})}{E},
    \label{eq:dIdV}
\end{align}
where $n_F(E)=1/(1+e^{E/k_BT})$ is the Fermi function. The ZBC is obtained by setting $\widetilde{V}=0$. 

\begin{figure}[b!]
    \centering
    \includegraphics[scale=0.4]{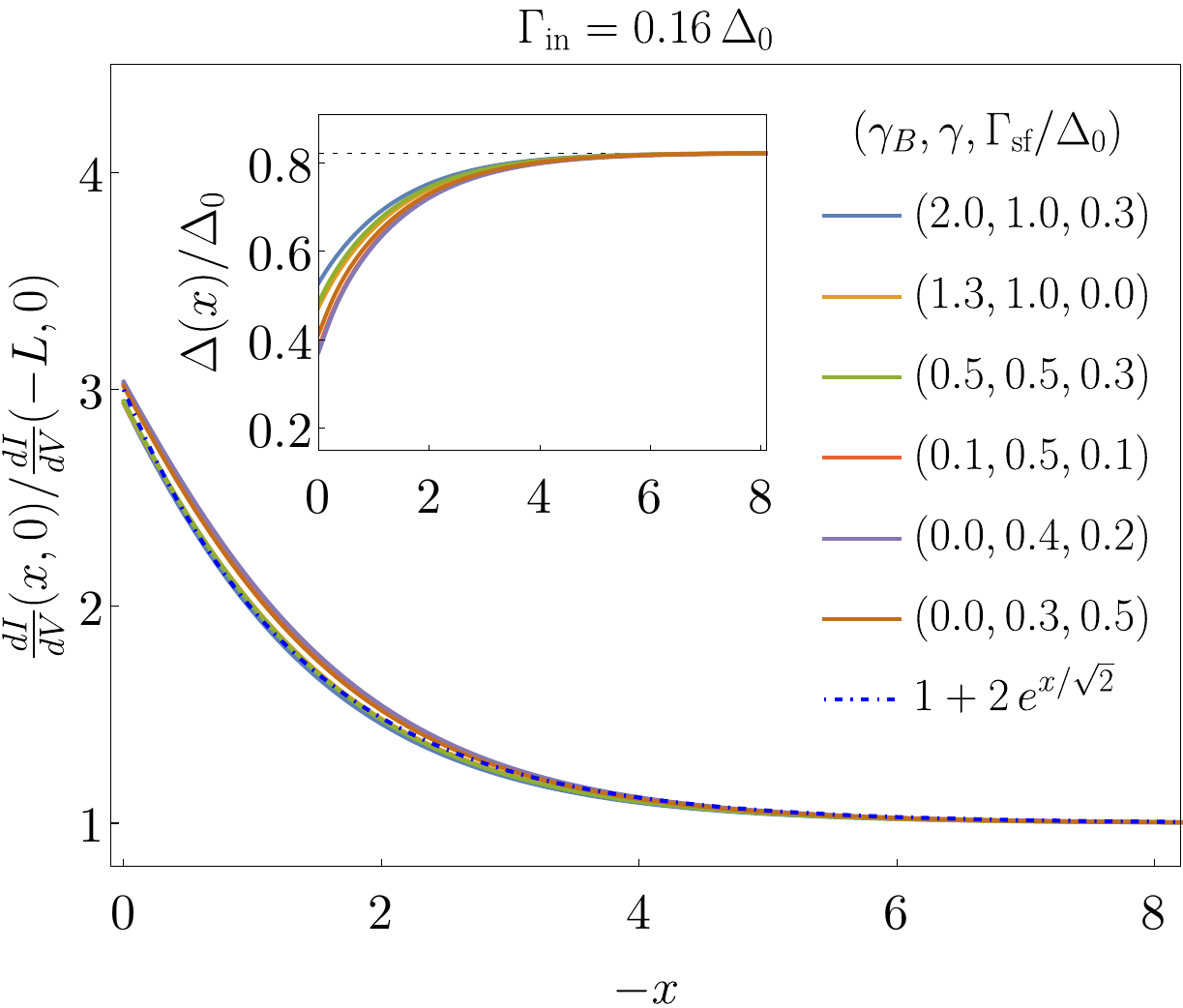}
    \caption{Zero bias conductance (ZBC) on the 1H side, normalized by its bulk value, for a variety of parameters, but fixing $\gin=0.16\D_0$ . With this choice of $\gin$ the bulk value is about $0.2056 N_S$. We have used $\gsf\Theta(x)$ as the spin-flip scattering term. The dot-dashed line shows the exponential rise in the ZBC that we wish to reproduce (see Fig.~\ref{fig:exp}). We have used $T=0.38$K and $\D_0=0.44\,$meV. Note that the $\sqrt{2}$ factor appears in the exponential because $x$ is in units of $\xi_S$ and the superconducting coherence length $\xi=\sqrt{2}\xi_S$. The inset shows the self-consistent gap $\D(x)$.}
    \label{fig:dIdV0}
\end{figure}

\begin{figure*}[t!]
    \centering
    \includegraphics[scale=0.4]{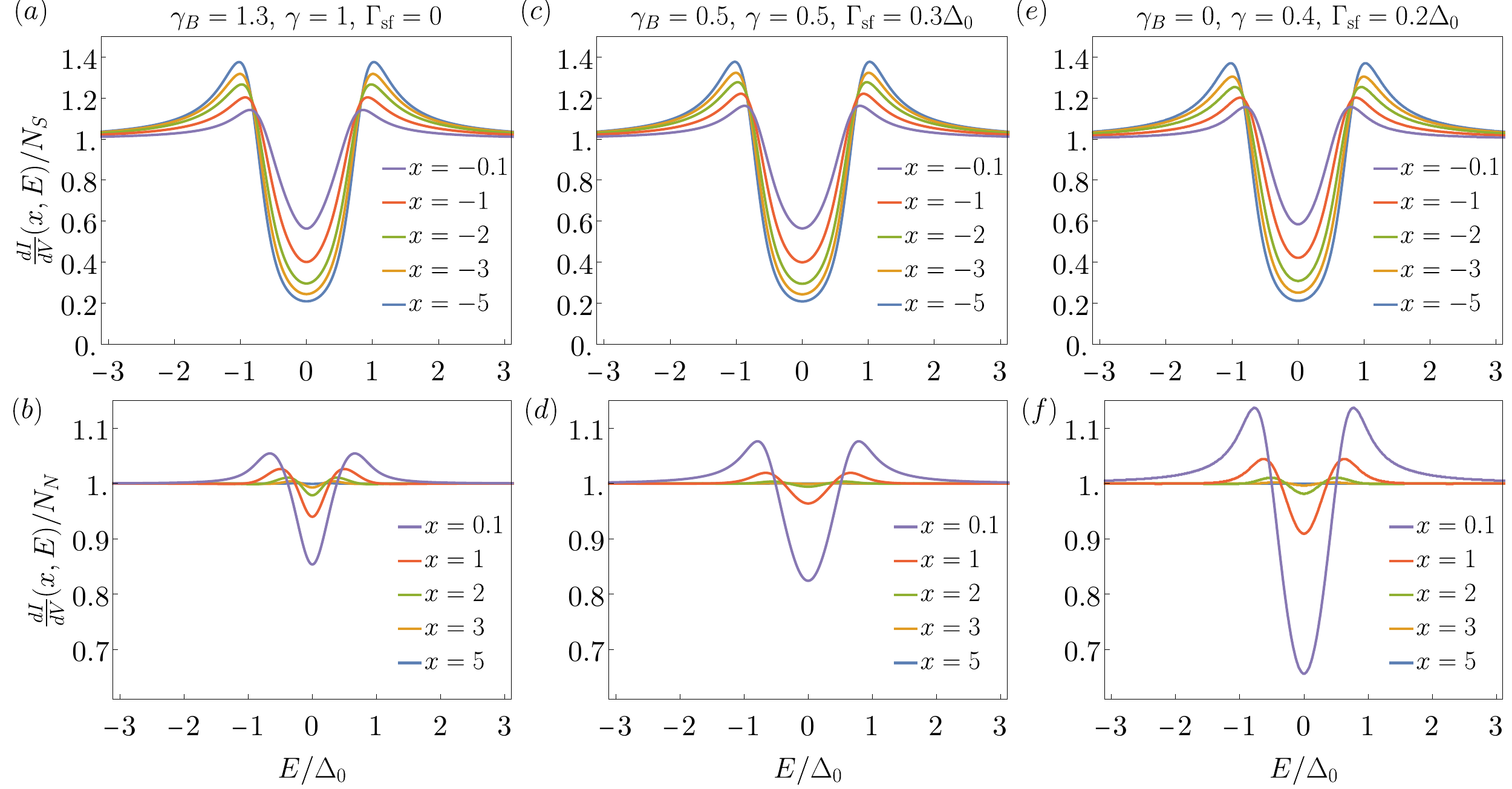}
    \caption{$dI/dV$ spectra on the 1H side (upper) and 1T side (lower), normalized by its normal-state DOS at Fermi energy, at various distances from the interface. We have taken $\gin=0.16\D_0$ and $T=0.38$K.}
    \label{fig:dIdV}
\end{figure*}

\begin{figure}[b!]
    \centering
    \includegraphics[scale=0.4]{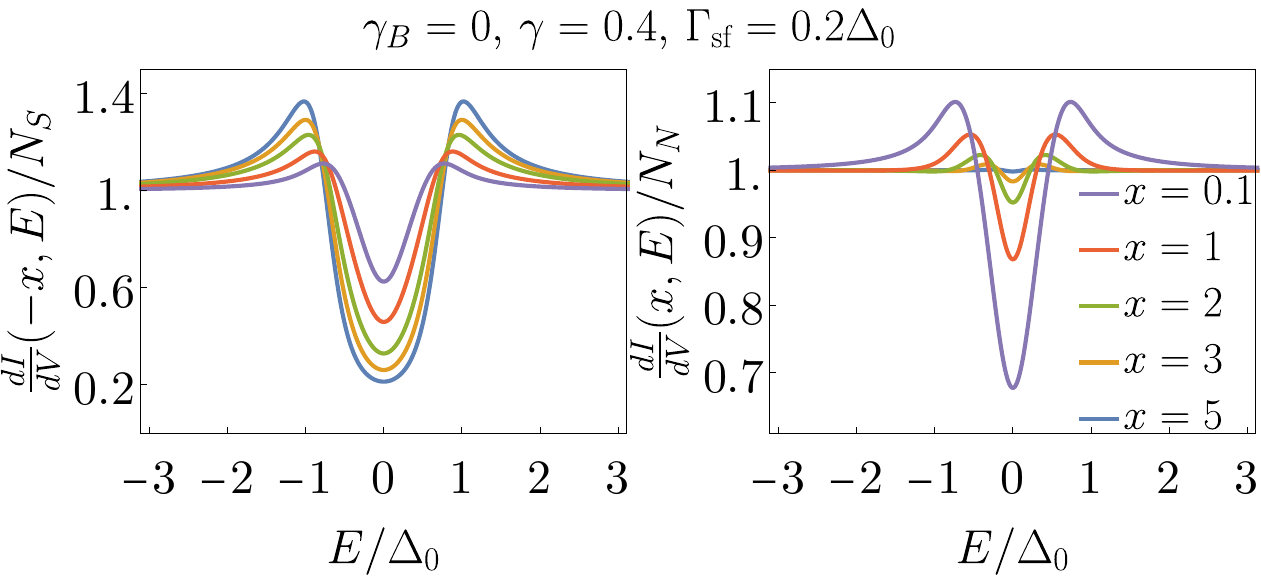}
    \caption{$dI/dV$ spectra on the 1H (left) and 1T (right) sides for $\gsf(x)=\gsf\,e^{-4x^2}$. We have used $\gin=0.16\D_0$ and $T=0.38$K.}
    \label{fig:dIdV_gauss}
\end{figure}

We find that various choices of $\gamma_B$, $\gamma$ and $\gsf$ produce the reported three-fold exponential rise in the ZBC on the 1H side with a localization length close to the 1H coherence length $\xi=\sqrt{\hbar D_S/\D_0}=\sqrt{2}\xi_S$. Some of these values are given in Table~\ref{table:4HbTaS2} and the corresponding ZBC plots are shown in Fig.~\ref{fig:dIdV0}. These curves compare well with the data shown in Fig.~\ref{fig:exp}. For a select few cases, we have shown the $dI/dV$ spectra at various distances from the interface in Fig.~\ref{fig:dIdV}. The bulk superconducting gap on the 1H side reduces to about $0.823\D_0$ because of the inelastic scattering term. Since we have taken a step like spin-flip scattering term, the $dI/dV$ spectra on the 1H side is predominantly determined by the inelastic scattering rate and temperature. It shows little change between different $(\gamma_B,\gamma,\gsf)$ values from Table~\ref{table:4HbTaS2} whereas the $dI/dV$ spectra on the 1T side is more sensitive to these parameters. The latter fact can be used to narrow down which set of values better explains the experiment by obtaining more data on the 1T side of the 1H step edge. The set of curves shown in Fig.~\ref{fig:dIdV} are in reasonable agreement with the curves shown in the inset in Fig.~\ref{fig:exp}.
We should point out that it is also 
possible to achieve a good agreement with the detailed shape of STM spectra by
using a localised $\gsf(x)$, an example of this is shown in Fig.~\ref{fig:dIdV_gauss}.  Thus, even the combination of the ZBC, $\Delta(x)$ around the interface, and the STM spectra does not provide strong constraints on the parameters appearing in the Usadel equations for a good agreement with experiments.

% --------------------------------------------------------------------------
\section{Conclusions}
\label{sec:conclusion}

We have presented a detailed study of the influence of the spatial profile of spin-flip scattering on the ZBC and LDOS in diffusive SN junctions. Our work provides non-self-consistent analytical solutions for the linearized Usadel equations, focusing on a step-like spin-flip scattering rate $\gsf\Theta(x)$. We have elucidated the roles of the mismatch parameter $\gamma$, the strength of the spin-flip term $\gsf$, and the interface transparency $\gamma_B$ in these systems. Our analysis reveals that increasing $\gamma$ and $\gsf$ leads to a more pronounced increase in the ZBC as one approaches the interface from the superconducting side, while $\gamma_B$ has the opposite effect. We have provided self-consistent calculations to corroborate these findings. Additionally, we have looked at the effects of having localized spin-flip scattering near the interface, offering further insights into the behavior of these junctions for various $\gsf(x)$.

Furthermore, we have applied our model to the specific case of 4Hb-TaS$_2$, providing a possible explanation for the observed rise in ZBC in this material. Although our self-consistent solution of the Usadel equations provide a good match for the ZBC and $dI/dV$ spectra, it is important to note that our one-dimensional modeling of the 1H step is highly simplified, leaving ample room for more sophisticated modeling and further exploration into the mechanisms responsible for these observations.  For example, it does not predict the different decay lengths of the ZBC away from step edges of different crystallographic orientations.

% --------------------------------------------------------------------------
\section{Acknowledgements}
% --------------------------------------------------------------------------
We are grateful to Haim Beidenkopf, Binghai Yan, Yuval Oreg, and Jonathan Ruhman for prior collaborations and discussions on 4Hb-TaS$_2$. We are especially grateful to Nurit Avraham who helped inspire and initiate this project in addition to participating in important discussions along the way. G.A.F. acknowledges funding from the National Science Foundation through DMR-2114825. G.A.F. acknowledges additional support the Alexander von Humboldt Foundation. PAL acknowledge support by DOE (USA) office of Basic Sciences Grant No. DE-FG02-03ER46076.  

% --------------------------------------------------------------------------
\appendix
% --------------------------------------------------------------------------
\section{Self-consistent solution for a uniform superconductor}
\label{appendix:selfc_uniform}
\begin{figure*}[ht]
    \centering
    \includegraphics[trim={0.2cm 0 0 0},clip,scale=0.34]{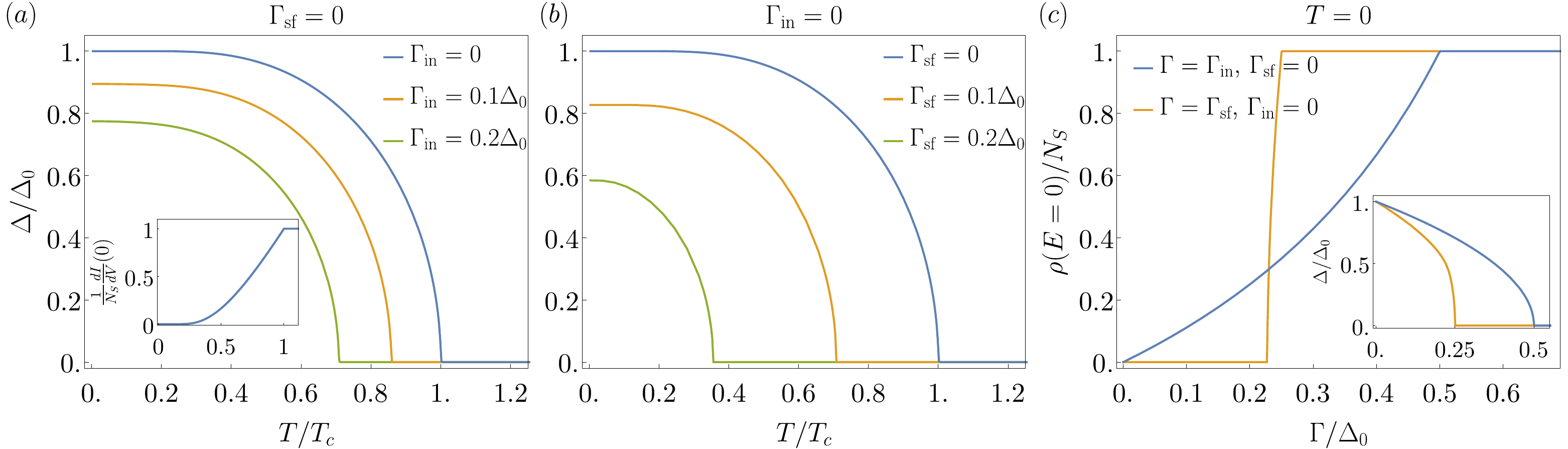}
    \caption{Self-consistent results for a uniform superconductor. Temperature dependence of $\D$ in the presence of (a) inelastic scattering, and (b) spin-flip scattering. The ZBC (calculated using Eq.~\ref{eq:dIdV}) for $\gin=\gsf=0$ is shown in the inset in (a). (c) Dependence of ZBC on $\gin$ (blue) and $\gsf$ (orange) at $T=0$. This was obtained using the corresponding $\D(\Gamma)$ curves shown in the inset. We used $E_c=100\D_0$ which gave $N_SV_S=0.18874$.}
    \label{fig:uniform_1}
\end{figure*}
For a uniform superconductor, the second-order spatial derivative in Eq.~\eqref{eq:Usadel_full} drops out and we get 
\begin{align}
\left(E+i\gin+2i\gsf\cosh{\theta}\right)\sinh{\theta}=\D\cosh{\theta}.
\label{eq:uniform_sc_full}
\end{align}
This needs to be solved together with the self-consistency condition given in Eq.~\eqref{eq:gap_full}, which becomes
\begin{align}
\D &= N_SV_S \int_0^{E_c} \dd{E} \, \tanh(\frac{E}{2k_BT})\text{Re}\left\{\sinh{\theta}\right\}.
\end{align}
Let us look at the case when $\gin=\gsf=0$. At $T=0$, $\D_0$ must be the self-consistent gap parameter. Since $\theta=\tanh^{-1}(\D_0/E)$, we can use this to determine $N_SV_S$ from
\begin{align}
 \frac{1}{N_SV_S} = \int_{\D_0}^{E_c}\frac{\dd{E}}{\sqrt{E^2-\D_0^2}} = \ln\left(\frac{E_c}{\D_0}+\sqrt{\frac{E_c^2}{\D_0^2}-1}\,\right),
\end{align}
which differs from the BCS result only in the negative sign inside both square roots. For $E_c=100\D_0$, this yields $N_SV_S=0.18874$ (used throughout this paper). We can now find and compare the temperature dependence of $\D$ with and without pair breaking terms. The plots are shown in Fig.~\ref{fig:uniform_1}(a) and (b), which makes it evident that spin-flip scattering suppresses superconductivity more strongly than the inelastic scattering. This fact is also captured in the inset in Fig.~\ref{fig:uniform_1}(c).

A simple calculation of how large the pair-breaking terms should be to destroy superconductivity at $T=0$ can be done by noting that as $\D$ becomes tiny, Eq.~\eqref{eq:uniform_sc_full} gives $\theta\approx \D/(E+i(\gin+2\gsf))$. Putting this in the self-consistency condition we get
\begin{align}
    \frac{1}{N_SV_S} &= \int_0^{E_c} \dd{E}\, \text{Re}\left\{\frac{1}{E+i(\gin+2\gsf)_c}\right\},
\end{align}
where $(\gin+2\gsf)_c$ refers to the critical value. Calculating further we get
\begin{align}
    &\ln\left(\frac{E_c}{\D_0}+\sqrt{\frac{E_c^2}{\D_0^2}-1}\,\right) = \int_0^{E_c} \dd{E}\, \frac{E}{E^2+(\gin+2\gsf)_c^2}, \\
    &\implies \frac{E_c}{\D_0}+\sqrt{\frac{E_c^2}{\D_0^2}-1} = \sqrt{\frac{E_c^2+(\gin+2\gsf)_c^2}{(\gin+2\gsf)_c^2}}, \\
    &\implies (\gin+2\gsf)_c \approx \frac{\D_0}{2}.
\end{align}
In the last step we used $E_c\gg\D_0$. This result also allows us to get $\Gamma_{\text{in},c}\approx\D_0/2$ when $\gsf=0$ and $\Gamma_{\text{sf},c}\approx\D_0/4$ when $\gin=0$ (see inset in Fig.~\ref{fig:uniform_1}(c)).

Let us also look at the ZBC for a uniform superconductor at $T=0$. Let us assume that $\D$ is the case specific self-consistent superconducting gap in the following. For $0<\gin<\Gamma_{\text{in},c}, \gsf=0$, $\theta(E=0)=\tanh^{-1}(\tfrac{\D}{i\gin})$ which gives a ZBC of $\tfrac{N_S}{\sqrt{1+(\D/\gin)^2}}$. For $\gin\geq\Gamma_{\text{in},c}$, we get the normal-state DOS $N_S$ at all energies. When $\gin=0, 0<\gsf<\Gamma_{\text{sf},c}$, $\theta(E=0)=\sinh^{-1}(\tfrac{\D}{2i\gsf})$ which gives a ZBC of $N_S\sqrt{1-(\tfrac{\D}{2\gsf})^2}\,\Theta(2\gsf-\D)$. We found that the equality $\D=2\gsf$ occurs at $\gsf\approx 0.22797\D_0$ for $E_c\gg\D_0$. Again, for $\gsf\geq\Gamma_{\text{sf},c}$, we get the normal-state DOS $N_S$ at all energies. These properties are shown in Fig.~\ref{fig:uniform_1}(c)

% --------------------------------------------------------------------------
\section{Self-consistent solution for SN junctions}
\label{appendix:selfc}
Expressing $x$ in units of $\xi_S$ and $\xi_N$ on the S and N sides, respectively, the Usadel equation in Eq.~\eqref{eq:Usadel_full} becomes   
\begin{align}
\begin{split}
\D_0 \pdv[2]{\theta_S}{x} + \left[iE-\gin-2\gsf(x)\cosh{\theta_S}\right]\sinh{\theta_S}\\
 - i\D(x)\cosh{\theta_S}=0,
\end{split}\\
\begin{split}
\D_0 \pdv[2]{\theta_N}{x} + \left[iE-\gin-2\gsf(x)\cosh{\theta_N}\right]\sinh{\theta_N}=0,
\end{split}
\end{align}
where $\theta_S=\theta(x<0)$ and $\theta_N=\theta(x>0)$. We will take $-L\leq x\leq L$ and use $L=100$. To solve this second-order nonlinear differential equation, we set it up as a system of first-order differential equations,
\begin{align}
 \pdv{\theta_S}{x} &= \phi_S, \\
 \begin{split}
 \pdv{\phi_S}{x} &= -\frac{1}{\D_0}\Big(\left[iE-\gin-2\gsf(x)\cosh{\theta_S}\right]\sinh{\theta_S},\\
 &\hspace{4cm} - i\D(x)\cosh{\theta_S}\Big),
 \end{split}\\
 \pdv{\theta_N}{x} &= \phi_N, \\
 \begin{split}
 \pdv{\phi_N}{x} &= -\frac{1}{\D_0}\left(iE-\gin-2\gsf(x)\cosh{\theta_N}\right)\sinh{\theta_N}.
\end{split}
\end{align}
The problem has boundary conditions both at the two ends and at the SN interface ($x=0$). This falls under the category of multipoint boundary value problems (BVPs). One way to solve this is to convert it into the usual two-point BVP. To do this, we perform a transformation on the independent variable, $x\rightarrow -x$, on the N side to fold it onto the S side. This also transforms $\theta_N(x,E) \rightarrow \widetilde{\theta}_N(x,E) = \theta_N(-x,E)$, and $\phi_N(x,E) \rightarrow \widetilde{\phi}_N(x,E) = \phi_N(-x,E)$. We now need to solve a new set of equations but over $[-L,0]$. This turns the condition at the SN interface into a boundary condition at the right end of the two-point BVP,
\begin{align}
 \pdv{\theta_S}{x} &= \phi_S, \\
 \begin{split}
 \pdv{\phi_S}{x} &= -\frac{1}{\D_0}\Big(\left[iE-\gin-2\gsf(x)\cosh{\theta_S}\right]\sinh{\theta_S}\\
 &\hspace{4cm} - i\D(x)\cosh{\theta_S}\Big),
 \end{split}\\
 -\pdv{\widetilde{\theta}_N}{x} &= \widetilde{\phi}_N, \\
 \begin{split}
 -\pdv{\widetilde{\phi}_N}{x} &= -\frac{1}{\D_0}\left(iE-\gin-2\gsf(-x)\cosh{\widetilde{\theta}_N}\right)\sinh{\widetilde{\theta}_N},
\end{split}
\end{align}
with the boundary conditions:
\begin{align}
    \theta_S(-L,E) &= \tanh^{-1}\left[\frac{\D(-L)}{E+i\gin} \right], \\
    \widetilde{\theta}_N(-L,E) &= 0, \\
    \phi_S(0,E) &= \gamma\,\widetilde{\phi}_N(0,E), \\
    \widetilde{\phi}_N(0,E) &= \frac{1}{\gamma_B}\sinh(\widetilde{\theta}_N(0,E)-\theta_S(0,E)).
\end{align}
When $\gamma_B=0$, the last boundary condition becomes $\widetilde{\theta}_N(0,E)=\theta_S(0,E)$, whereas for $\gamma_B \rightarrow \infty$, it gives $\widetilde{\phi}_N(0,E)=0$.

We solve this two-point BVP numerically in Python using the scipy.integrate.solve\_bvp function, starting with
\begin{equation}
   \D(x) = \D_0.
\end{equation}
The the self-consistency condition gives the gap on the S side for the next iteration:
\begin{align}
\D(x) &= N_SV_S \int_0^{E_c} \dd{E} \, \tanh(\frac{E}{2k_BT}) \text{Re}\left\{\sinh{\theta_S}\right\}.
\end{align}
The procedure is repeated until convergence is achieved, typically within 30 iterations. As before, we use $\D_0=0.44\,$meV, $E_c=100\D_0$, and $N_SV_S=0.18874$.

% --------------------------------------------------------------------------
\section{Solution for linearized Usadel equations}
\label{appendix:nonselfc}

Let us take a step-wise constant gap across the interface,
\begin{equation}
   \D(x) =
    \begin{cases}
      \D_0 & \text{if $x<0$}\\
      0 & \text{if $x>0$}
    \end{cases}.
    \label{}
\end{equation}
With this choice, the Usadel equations on the two sides become,
\begin{equation}
    \D_0 \frac{\partial^2 \theta_S}{\partial x^2} +[iE-\gin] \sinh\theta_S-i\D_0\cosh\theta_S = 0,
\label{eq:theta_S}
\end{equation}
\begin{equation}
    \D_0 \frac{\partial^2 \theta_N}{\partial x^2} + [iE-\gin - 2\gsf\cosh\theta_N] \sinh\theta_N = 0,
\label{eq:theta_N}
\end{equation}
where $\theta_S=\theta(x<0)$ and $\theta_N=\theta(x>0)$. While the inelastic broadening $\gin$ is taken to be constant  throughout the system, we consider a step-like spin-flip scattering across the interface,
\begin{equation}
  \gsf(x) =
    \begin{cases}
      0 & \text{if $x<0$}\\
      \gsf & \text{if $x>0$}
    \end{cases}.
\end{equation}
We will assume $E$, $\D_0$, $\gin$, and $\gsf$ to be positive in the remainder of this section.

It is straightforward to linearize Eq.\eqref{eq:theta_N} to
\begin{equation}
\D_0 \frac{\partial^2 \theta_N}{\partial x^2} +[iE -\Gamma_{in}-2\gsf]\theta_N = 0,
\label{eq:theta_N_linearized}
\end{equation}
which is valid for a small $\theta_N$ corresponding to a small induced proximity gap. This has a solution of the form
\begin{equation}
\theta_N(x,E) = A(E)e^{-\omega x},
\label{eq:theta_N_step_soln}
\end{equation}
where $\omega=\sqrt{(-iE+\gin+2\gsf)/\D_0}$ and $A$ is an unknown function at this moment. We take the principal square root here so that $\omega$ has a positive real part.

On the superconducting side of the interface, the order parameter $\D(x)$ (and correspondingly $\theta(x,E)$) is not small and one would not expect to be able to expand the Usadel equation around $\theta(x,E)\approx0$, as was done on the normal side of the interface. However, an expansion around the bulk, uniform superconducting value given by $\theta_0(E) = \tanh^{-1} \left(\frac{\D_0}{E+i\gin}\right)$ should yield sensible results around the interface. We thus substitute
\begin{equation}
\theta_S(x,E) = \theta_0(E)+\delta\theta_S(x,E),
\end{equation}
into Eq.~\eqref{eq:theta_S} and using
\begin{eqnarray}
    \sinh(\theta_0+\delta \theta_S) \approx \sinh(\theta_0)+\cosh(\theta_0)\delta \theta_S, \\
    \cosh(\theta_0+\delta \theta_S) \approx \cosh(\theta_0)+\sinh(\theta_0)\delta \theta_S,
\end{eqnarray}
obtain (to linear order in $\delta\theta_S$)
\begin{equation}
\D_0 \frac{\partial^2 \delta\theta_S}{\partial x^2} +[iE-\gin] \cosh(\theta_0)\delta\theta_S - i\D_0\sinh(\theta_0)\delta \theta_S=0,
\end{equation}
which simplifies to
\begin{equation}
\D_0 \frac{\partial^2 \delta\theta_S}{\partial x^2} + i\sqrt{(E+i\gin)^2-\D_0^2}\, \delta\theta_S = 0.
\label{eq:theta_S_linearized_Delta}
\end{equation}
Similar to the N side, it has a solution of the form
\begin{equation}
\delta\theta_S(x,E) = B(E)e^{\omega' x},
\label{eq:theta_S_step_soln}
\end{equation}
where $\omega'=\sqrt{-i\sqrt{(E+i\gin)^2/\D_0^2-1}}$ (we take the principal square root at both levels so that $\omega'$ is guaranteed to have a positive real part) and $B$ is another unknown function. We determine $A$ and $B$ from the condition at the interface.

Assuming perfect transparency ($\gamma_B=0$), the condition on $\theta$ at the interface gives 
\begin{equation}
    A=\theta_0+B.
\end{equation}
Similarly, the condition on its derivative gives
\begin{equation}
    -\gamma \omega A= \omega' B.
\end{equation}
We solve for $A$ and $B$ to get:
\begin{align}
    A &= \frac{\theta_0}{1+\gamma\omega/\omega'} \\
    B &= \frac{-\theta_0}{1+\omega'/\gamma\omega}.
\end{align}
Thus, we obtain the following solution for the linearized Usadel equation
\begin{align}
    \theta_S(x,E) &= \theta_0 - \frac{\theta_0}{1+\omega'/\gamma\omega} \,e^{\omega'x},\quad x\leq 0 \label{eq:theta_S_lin_soln}\\
    \theta_N(x,E) &= \frac{\theta_0}{1+\gamma\omega/\omega'} \,e^{-\omega x},\quad x\geq 0. \label{eq:theta_N_lin_soln}
\end{align}
\vspace{0.1cm}
It is straightforward to obtain the LDOS from this as,
\begin{align}
\rho(x,E) =
    \begin{cases}
      N_S \, \text{Re}\{\cosh(\theta_S)\} & \text{if $x<0$}\\
      N_N \, \text{Re}\{\cosh(\theta_N)\} & \text{if $x>0$}
    \end{cases}.
\end{align}

% -------------------------------------
\subsection{LDOS for $E\ll\D_0$ when $\gin=\gsf=0$}

In absence of any pair-breaking scattering mechanisms, the LDOS near $E=0$ is given by
\begin{align}
\rho(x,E) \approx
    \begin{cases}
        \frac{N_S \pi}{2\sqrt{2}}\gamma\,e^x\sqrt{\frac{|E|}{\D_0}} ,& \text{$x<0$}\\
        \frac{N_N \pi}{2\sqrt{2}}(\gamma+x)\sqrt{\frac{|E|}{\D_0}} ,& \text{$x>0$}
    \end{cases},
\end{align}
and shows a $\sqrt{|E|}$ dependence. Note that if one uses the non-self-consistent analytical solution from~\cite{Belzig:prb96}, the factor $\frac{\pi}{2\sqrt{2}}$ drops out.

% -------------------------------------
\subsection{LDOS for $E\ll\D_0$ when $\gin+\gsf\neq 0$}

When either of $\gin$, $\gsf$ is nonzero ($\gin+\gsf\neq 0$), the LDOS is given by
\begin{widetext}
\begin{align}
\rho(x,E) \approx
    \begin{cases}
        N_S \cos(\tan^{-1}(\frac{\D_0}{\gin})\frac{1+\gamma\sqrt{\frac{\gin+2\gsf}{\sqrt{\gin^2+\D_0^2}}}\left(1-e^{x\sqrt{\sqrt{1+\gin^2/\D_0^2}}}\right)}{1+\gamma\sqrt{\frac{\gin+2\gsf}{\sqrt{\gin^2+\D_0^2}}}})\Big(1+\mathcal{O}\left(\tfrac{E^2}{\D_0^2}\right)\Big) ,& \text{$x<0$}\\
        N_N \cos(\tan^{-1}(\frac{\D_0}{\gin})\frac{e^{-x\sqrt{(\gin+2\gsf)/\D_0}}}{1+\gamma\sqrt{\frac{\gin+2\gsf}{\sqrt{\gin^2+\D_0^2}}}})\Big(1+\mathcal{O}\left(\tfrac{E^2}{\D_0^2}\right)\Big) ,& \text{$x>0$}
    \end{cases},
\end{align}
\end{widetext}
which has a finite value at $E=0$ and follows a $E^2$ dependence.
Let us explicitly calculate the ZBC on the N and S sides of the interface. For the N side we get
\begin{align}
\begin{split}
    & \frac{\rho(x>0,0)}{N_N} = \cos \left[\tfrac{\tan ^{-1}\left(\frac{\D_0 }{\gin}\right) e^{-x \sqrt{\frac{\gin+2\gsf}{\D_0}}}}{1+\gamma\sqrt{\frac{\gin+2\gsf}{\sqrt{\gin^2+\D_0^2}}}}\right],
\end{split}
\end{align}
which has a bulk value of $\rho(+\infty,0) = N_N$, as expected. Similarly, the ZBC on S side is given by
\begin{align}
\begin{split}
    & \frac{\rho(x<0,0)}{N_S} = \cos \left[\tan ^{-1}\left(\tfrac{\D_0 }{\gin}\right)\left(1-\tfrac{e^{x \sqrt{\sqrt{1+\gin^2/\D_0^2}}}}{1+\frac{1}{\gamma}\sqrt{\frac{\sqrt{\gin^2+\D_0^2}}{\gin+2\gsf}}}\right)\right]
\end{split}
\end{align}
and has the bulk value
\begin{equation}
    \rho(-\infty,0) = \frac{N_S}{\sqrt{1+(\D_0/\gin)^2}}.
\end{equation}
At the interface, we find that
\begin{align}
    \frac{\rho(0^-,0)}{N_S} = \frac{\rho(0^+,0)}{N_N} = \cos \left[\frac{\tan ^{-1}\left(\frac{\D_0 }{\gin}\right)}{1+\gamma\sqrt{\frac{\gin+2\gsf}{\sqrt{\gin^2+\D_0^2}}}}\right],
\end{align}
which is always nonzero for $\gin+\gsf\neq 0$. 

% -------------------------------------
\subsection{Solution for $\gamma_B\neq 0$}

When $\gamma_B\neq 0$, the boundary conditions give
\begin{equation}
    \omega' B = -\gamma \omega A = \frac{\gamma}{\gamma_B}\sinh(A-B-\theta_0)
\end{equation}
and we obtain
\begin{align}
    \theta_S(x,E) &= \theta_0 - \frac{\gamma\omega A}{\omega'} \,e^{\omega'x},\quad x<0 \label{eq:theta_S_lin_soln2}\\
    \theta_N(x,E) &= A \,e^{-\omega x},\quad x>0 \label{eq:theta_N_lin_soln2}
\end{align}
where, $A(E)$ needs to be found numerically for an arbitrary value of $\gamma_B$ by solving
\begin{align}
    \gamma_B \omega A = \sinh(\theta_0 - A\left(1+\frac{\gamma \omega}{\omega'}\right)).\label{eq:solveA}
\end{align}
However, we can get an approximate solution for $A$ in two extreme limits. Let us first consider the case $\gamma_B\ll 1$. This condition allows us to simplify Eq.~\eqref{eq:solveA} to get
\begin{align}
    A \approx \frac{\theta_0}{1+\gamma\omega/\omega' + \gamma_B\omega}.
\end{align}
From this, we calculate the ZBC
\begin{align}
\begin{split}
    & \frac{\rho(x>0,0)}{N_N} = \cos \left[\tfrac{\tan ^{-1}\left(\frac{\D_0 }{\gin}\right) e^{-x \sqrt{\frac{\gin+2\gsf}{\D_0}}}}{1+\gamma\sqrt{\frac{\gin+2\gsf}{\sqrt{\gin^2+\D_0^2}}}+\gamma_B\sqrt{\frac{\gin+2\gsf}{\D_0}}}\right],
\end{split} \\
\begin{split}
    & \frac{\rho(x<0,0)}{N_S} = \cos \left[\tan ^{-1}\left(\tfrac{\D_0 }{\gin}\right) \vphantom{\Bigg|}\right.\\
    &\hspace{2.2cm} \left. -\tfrac{\tan ^{-1}\left(\tfrac{\D_0 }{\gin}\right)e^{x \sqrt{\sqrt{1+\gin^2/\D_0^2}}}}{1+\frac{1}{\gamma}\sqrt{\frac{\sqrt{\gin^2+\D_0^2}}{\gin+2\gsf}}+\frac{\gamma_B}{\gamma}\sqrt{\frac{\sqrt{\gin^2+\D_0^2}}{\D_0}}}\right].
\end{split}
\end{align}
\vspace{0.1cm}
Similarly, when $\gamma_B\gg 1$, Eq.~\eqref{eq:solveA} simplifies to
\begin{align}
     A \approx \frac{\tanh(\theta_0)}{1+\gamma\omega/\omega' + i\gamma_B\omega\omega'^2\tanh(\theta_0)} \approx \frac{1}{i\gamma_B\omega\omega'^2}
\end{align}
which gives the ZBC
\begin{align}
\begin{split}
    & \frac{\rho(x>0,0)}{N_N} = \cos \left[\tfrac{ e^{-x \sqrt{\frac{\gin+2\gsf}{\D_0}}}}{\gamma_B\sqrt{\frac{\gin+2\gsf}{\D_0}}\frac{\sqrt{\gin^2+\D_0^2}}{\D_0}}\right],
\end{split} \\
\begin{split}
    & \frac{\rho(x<0,0)}{N_S} = \cos \left[\tan ^{-1}\left(\tfrac{\D_0 }{\gin}\right)-\tfrac{e^{x \sqrt{\sqrt{1+\gin^2/\D_0^2}}}}{\frac{\gamma_B}{\gamma}\left(1+\frac{\gin^2}{\D_0^2}\right)^{3/4}}\right]
\end{split}.
\end{align}
In both limits, an increase in $\gamma_B$ leads to decrease in $\rho(0^-,0)$ and increase in $\rho(0^+,0)$. Also note that for a nonzero $\gamma_B$,
\begin{align}
    \frac{\rho(0^-,0)}{N_S} \neq \frac{\rho(0^+,0)}{N_N}.
\end{align}
\vspace{1cm}

\bibliography{UsadelRef}

\begin{thebibliography}{32}%
\makeatletter
\providecommand \@ifxundefined [1]{%
 \@ifx{#1\undefined}
}%
\providecommand \@ifnum [1]{%
 \ifnum #1\expandafter \@firstoftwo
 \else \expandafter \@secondoftwo
 \fi
}%
\providecommand \@ifx [1]{%
 \ifx #1\expandafter \@firstoftwo
 \else \expandafter \@secondoftwo
 \fi
}%
\providecommand \natexlab [1]{#1}%
\providecommand \enquote  [1]{``#1''}%
\providecommand \bibnamefont  [1]{#1}%
\providecommand \bibfnamefont [1]{#1}%
\providecommand \citenamefont [1]{#1}%
\providecommand \href@noop [0]{\@secondoftwo}%
\providecommand \href [0]{\begingroup \@sanitize@url \@href}%
\providecommand \@href[1]{\@@startlink{#1}\@@href}%
\providecommand \@@href[1]{\endgroup#1\@@endlink}%
\providecommand \@sanitize@url [0]{\catcode `\\12\catcode `\$12\catcode
  `\&12\catcode `\#12\catcode `\^12\catcode `\_12\catcode `\%12\relax}%
\providecommand \@@startlink[1]{}%
\providecommand \@@endlink[0]{}%
\providecommand \url  [0]{\begingroup\@sanitize@url \@url }%
\providecommand \@url [1]{\endgroup\@href {#1}{\urlprefix }}%
\providecommand \urlprefix  [0]{URL }%
\providecommand \Eprint [0]{\href }%
\providecommand \doibase [0]{https://doi.org/}%
\providecommand \selectlanguage [0]{\@gobble}%
\providecommand \bibinfo  [0]{\@secondoftwo}%
\providecommand \bibfield  [0]{\@secondoftwo}%
\providecommand \translation [1]{[#1]}%
\providecommand \BibitemOpen [0]{}%
\providecommand \bibitemStop [0]{}%
\providecommand \bibitemNoStop [0]{.\EOS\space}%
\providecommand \EOS [0]{\spacefactor3000\relax}%
\providecommand \BibitemShut  [1]{\csname bibitem#1\endcsname}%
\let\auto@bib@innerbib\@empty
%</preamble>
\bibitem [{\citenamefont {De~Gennes}(1964)}]{deGennes:rmp64}%
  \BibitemOpen
  \bibfield  {author} {\bibinfo {author} {\bibfnamefont {P.~G.}\ \bibnamefont
  {De~Gennes}},\ }\bibfield  {title} {\bibinfo {title} {Boundary effects in
  superconductors},\ }\href {https://doi.org/10.1103/RevModPhys.36.225}
  {\bibfield  {journal} {\bibinfo  {journal} {Rev. Mod. Phys.}\ }\textbf
  {\bibinfo {volume} {36}},\ \bibinfo {pages} {225} (\bibinfo {year}
  {1964})}\BibitemShut {NoStop}%
\bibitem [{\citenamefont {Pannetier}\ and\ \citenamefont
  {Courtois}(2000)}]{pannetier2000}%
  \BibitemOpen
  \bibfield  {author} {\bibinfo {author} {\bibfnamefont {B.}~\bibnamefont
  {Pannetier}}\ and\ \bibinfo {author} {\bibfnamefont {H.}~\bibnamefont
  {Courtois}},\ }\bibfield  {title} {\bibinfo {title} {Andreev reflection and
  proximity effect},\ }\href@noop {} {\bibfield  {journal} {\bibinfo  {journal}
  {Journal of low temperature physics}\ }\textbf {\bibinfo {volume} {118}},\
  \bibinfo {pages} {599} (\bibinfo {year} {2000})}\BibitemShut {NoStop}%
\bibitem [{\citenamefont {Gu\'eron}\ \emph {et~al.}(1996)\citenamefont
  {Gu\'eron}, \citenamefont {Pothier}, \citenamefont {Birge}, \citenamefont
  {Esteve},\ and\ \citenamefont {Devoret}}]{Gueron:prl96}%
  \BibitemOpen
  \bibfield  {author} {\bibinfo {author} {\bibfnamefont {S.}~\bibnamefont
  {Gu\'eron}}, \bibinfo {author} {\bibfnamefont {H.}~\bibnamefont {Pothier}},
  \bibinfo {author} {\bibfnamefont {N.~O.}\ \bibnamefont {Birge}}, \bibinfo
  {author} {\bibfnamefont {D.}~\bibnamefont {Esteve}},\ and\ \bibinfo {author}
  {\bibfnamefont {M.~H.}\ \bibnamefont {Devoret}},\ }\bibfield  {title}
  {\bibinfo {title} {Superconducting proximity effect probed on a mesoscopic
  length scale},\ }\href {https://doi.org/10.1103/PhysRevLett.77.3025}
  {\bibfield  {journal} {\bibinfo  {journal} {Phys. Rev. Lett.}\ }\textbf
  {\bibinfo {volume} {77}},\ \bibinfo {pages} {3025} (\bibinfo {year}
  {1996})}\BibitemShut {NoStop}%
\bibitem [{\citenamefont {Moussy}\ \emph {et~al.}(2001)\citenamefont {Moussy},
  \citenamefont {Courtois},\ and\ \citenamefont {Pannetier}}]{Moussy:EPL2001}%
  \BibitemOpen
  \bibfield  {author} {\bibinfo {author} {\bibfnamefont {N.}~\bibnamefont
  {Moussy}}, \bibinfo {author} {\bibfnamefont {H.}~\bibnamefont {Courtois}},\
  and\ \bibinfo {author} {\bibfnamefont {B.}~\bibnamefont {Pannetier}},\
  }\bibfield  {title} {\bibinfo {title} {Local spectroscopy of a proximity
  superconductor at very low temperature},\ }\href
  {https://iopscience.iop.org/article/10.1209/epl/i2001-00361-2} {\bibfield
  {journal} {\bibinfo  {journal} {Europhysics Letters ({EPL})}\ }\textbf
  {\bibinfo {volume} {55}},\ \bibinfo {pages} {861} (\bibinfo {year}
  {2001})}\BibitemShut {NoStop}%
\bibitem [{\citenamefont {Vinet}\ \emph {et~al.}(2001)\citenamefont {Vinet},
  \citenamefont {Chapelier},\ and\ \citenamefont {Lefloch}}]{vinet2001}%
  \BibitemOpen
  \bibfield  {author} {\bibinfo {author} {\bibfnamefont {M.}~\bibnamefont
  {Vinet}}, \bibinfo {author} {\bibfnamefont {C.}~\bibnamefont {Chapelier}},\
  and\ \bibinfo {author} {\bibfnamefont {F.}~\bibnamefont {Lefloch}},\
  }\bibfield  {title} {\bibinfo {title} {Spatially resolved spectroscopy on
  superconducting proximity nanostructures},\ }\href
  {https://doi.org/10.1103/PhysRevB.63.165420} {\bibfield  {journal} {\bibinfo
  {journal} {Phys. Rev. B}\ }\textbf {\bibinfo {volume} {63}},\ \bibinfo
  {pages} {165420} (\bibinfo {year} {2001})}\BibitemShut {NoStop}%
\bibitem [{\citenamefont {Meschke}\ \emph {et~al.}(2011)\citenamefont
  {Meschke}, \citenamefont {Peltonen}, \citenamefont {Pekola},\ and\
  \citenamefont {Giazotto}}]{meschke2011}%
  \BibitemOpen
  \bibfield  {author} {\bibinfo {author} {\bibfnamefont {M.}~\bibnamefont
  {Meschke}}, \bibinfo {author} {\bibfnamefont {J.~T.}\ \bibnamefont
  {Peltonen}}, \bibinfo {author} {\bibfnamefont {J.~P.}\ \bibnamefont
  {Pekola}},\ and\ \bibinfo {author} {\bibfnamefont {F.}~\bibnamefont
  {Giazotto}},\ }\bibfield  {title} {\bibinfo {title} {Tunnel spectroscopy of a
  proximity josephson junction},\ }\href
  {https://doi.org/10.1103/PhysRevB.84.214514} {\bibfield  {journal} {\bibinfo
  {journal} {Phys. Rev. B}\ }\textbf {\bibinfo {volume} {84}},\ \bibinfo
  {pages} {214514} (\bibinfo {year} {2011})}\BibitemShut {NoStop}%
\bibitem [{\citenamefont {Usadel}(1970)}]{Usadel:prl70}%
  \BibitemOpen
  \bibfield  {author} {\bibinfo {author} {\bibfnamefont {K.~D.}\ \bibnamefont
  {Usadel}},\ }\bibfield  {title} {\bibinfo {title} {Generalized diffusion
  equation for superconducting alloys},\ }\href
  {https://doi.org/10.1103/PhysRevLett.25.507} {\bibfield  {journal} {\bibinfo
  {journal} {Phys. Rev. Lett.}\ }\textbf {\bibinfo {volume} {25}},\ \bibinfo
  {pages} {507} (\bibinfo {year} {1970})}\BibitemShut {NoStop}%
\bibitem [{\citenamefont {Belzig}\ \emph {et~al.}(1996)\citenamefont {Belzig},
  \citenamefont {Bruder},\ and\ \citenamefont {Sch\"on}}]{Belzig:prb96}%
  \BibitemOpen
  \bibfield  {author} {\bibinfo {author} {\bibfnamefont {W.}~\bibnamefont
  {Belzig}}, \bibinfo {author} {\bibfnamefont {C.}~\bibnamefont {Bruder}},\
  and\ \bibinfo {author} {\bibfnamefont {G.}~\bibnamefont {Sch\"on}},\
  }\bibfield  {title} {\bibinfo {title} {Local density of states in a dirty
  normal metal connected to a superconductor},\ }\href
  {https://doi.org/10.1103/PhysRevB.54.9443} {\bibfield  {journal} {\bibinfo
  {journal} {Phys. Rev. B}\ }\textbf {\bibinfo {volume} {54}},\ \bibinfo
  {pages} {9443} (\bibinfo {year} {1996})}\BibitemShut {NoStop}%
\bibitem [{\citenamefont {Crouzy}\ \emph {et~al.}(2005)\citenamefont {Crouzy},
  \citenamefont {Bascones},\ and\ \citenamefont {Ivanov}}]{Crouzy:prb05}%
  \BibitemOpen
  \bibfield  {author} {\bibinfo {author} {\bibfnamefont {B.}~\bibnamefont
  {Crouzy}}, \bibinfo {author} {\bibfnamefont {E.}~\bibnamefont {Bascones}},\
  and\ \bibinfo {author} {\bibfnamefont {D.~A.}\ \bibnamefont {Ivanov}},\
  }\bibfield  {title} {\bibinfo {title} {Minigap in a superconductor--normal
  metal junction with paramagnetic impurities},\ }\href
  {https://doi.org/10.1103/PhysRevB.72.092501} {\bibfield  {journal} {\bibinfo
  {journal} {Phys. Rev. B}\ }\textbf {\bibinfo {volume} {72}},\ \bibinfo
  {pages} {092501} (\bibinfo {year} {2005})}\BibitemShut {NoStop}%
\bibitem [{\citenamefont {Golubov}\ \emph {et~al.}(1997)\citenamefont
  {Golubov}, \citenamefont {Wilhelm},\ and\ \citenamefont
  {Zaikin}}]{golubov1997}%
  \BibitemOpen
  \bibfield  {author} {\bibinfo {author} {\bibfnamefont {A.~A.}\ \bibnamefont
  {Golubov}}, \bibinfo {author} {\bibfnamefont {F.~K.}\ \bibnamefont
  {Wilhelm}},\ and\ \bibinfo {author} {\bibfnamefont {A.~D.}\ \bibnamefont
  {Zaikin}},\ }\bibfield  {title} {\bibinfo {title} {Coherent charge transport
  in metallic proximity structures},\ }\href
  {https://doi.org/10.1103/PhysRevB.55.1123} {\bibfield  {journal} {\bibinfo
  {journal} {Phys. Rev. B}\ }\textbf {\bibinfo {volume} {55}},\ \bibinfo
  {pages} {1123} (\bibinfo {year} {1997})}\BibitemShut {NoStop}%
\bibitem [{\citenamefont {Yip}(1995)}]{yip1995}%
  \BibitemOpen
  \bibfield  {author} {\bibinfo {author} {\bibfnamefont {S.}~\bibnamefont
  {Yip}},\ }\bibfield  {title} {\bibinfo {title} {Conductance anomalies for
  normal-metal--insulator--superconductor contacts},\ }\href
  {https://doi.org/10.1103/PhysRevB.52.15504} {\bibfield  {journal} {\bibinfo
  {journal} {Phys. Rev. B}\ }\textbf {\bibinfo {volume} {52}},\ \bibinfo
  {pages} {15504} (\bibinfo {year} {1995})}\BibitemShut {NoStop}%
\bibitem [{\citenamefont {Nayak}\ \emph {et~al.}(2021)\citenamefont {Nayak},
  \citenamefont {Steinbok}, \citenamefont {Roet}, \citenamefont {Koo},
  \citenamefont {Margalit}, \citenamefont {Feldman}, \citenamefont {Almoalem},
  \citenamefont {Kanigel}, \citenamefont {Fiete}, \citenamefont {Yan} \emph
  {et~al.}}]{nayak2021}%
  \BibitemOpen
  \bibfield  {author} {\bibinfo {author} {\bibfnamefont {A.~K.}\ \bibnamefont
  {Nayak}}, \bibinfo {author} {\bibfnamefont {A.}~\bibnamefont {Steinbok}},
  \bibinfo {author} {\bibfnamefont {Y.}~\bibnamefont {Roet}}, \bibinfo {author}
  {\bibfnamefont {J.}~\bibnamefont {Koo}}, \bibinfo {author} {\bibfnamefont
  {G.}~\bibnamefont {Margalit}}, \bibinfo {author} {\bibfnamefont
  {I.}~\bibnamefont {Feldman}}, \bibinfo {author} {\bibfnamefont
  {A.}~\bibnamefont {Almoalem}}, \bibinfo {author} {\bibfnamefont
  {A.}~\bibnamefont {Kanigel}}, \bibinfo {author} {\bibfnamefont {G.~A.}\
  \bibnamefont {Fiete}}, \bibinfo {author} {\bibfnamefont {B.}~\bibnamefont
  {Yan}}, \emph {et~al.},\ }\bibfield  {title} {\bibinfo {title} {Evidence of
  topological boundary modes with topological nodal-point superconductivity},\
  }\href {https://doi.org/10.1038/s41567-021-01376-z} {\bibfield  {journal}
  {\bibinfo  {journal} {Nature physics}\ }\textbf {\bibinfo {volume} {17}},\
  \bibinfo {pages} {1413} (\bibinfo {year} {2021})}\BibitemShut {NoStop}%
\bibitem [{\citenamefont {Raj}\ \emph {et~al.}(2024)\citenamefont {Raj},
  \citenamefont {Postlewaite}, \citenamefont {Chaudhary},\ and\ \citenamefont
  {Fiete}}]{raj2023nonlinear}%
  \BibitemOpen
  \bibfield  {author} {\bibinfo {author} {\bibfnamefont {A.}~\bibnamefont
  {Raj}}, \bibinfo {author} {\bibfnamefont {A.}~\bibnamefont {Postlewaite}},
  \bibinfo {author} {\bibfnamefont {S.}~\bibnamefont {Chaudhary}},\ and\
  \bibinfo {author} {\bibfnamefont {G.~A.}\ \bibnamefont {Fiete}},\ }\bibfield
  {title} {\bibinfo {title} {Nonlinear optical responses in multiorbital
  topological superconductors},\ }\href
  {https://doi.org/10.1103/PhysRevB.109.184514} {\bibfield  {journal} {\bibinfo
   {journal} {Phys. Rev. B}\ }\textbf {\bibinfo {volume} {109}},\ \bibinfo
  {pages} {184514} (\bibinfo {year} {2024})}\BibitemShut {NoStop}%
\bibitem [{\citenamefont {Ribak}\ \emph {et~al.}(2020)\citenamefont {Ribak},
  \citenamefont {Skiff}, \citenamefont {Mograbi}, \citenamefont {Rout},
  \citenamefont {Fischer}, \citenamefont {Ruhman}, \citenamefont {Chashka},
  \citenamefont {Dagan},\ and\ \citenamefont {Kanigel}}]{ribak2020}%
  \BibitemOpen
  \bibfield  {author} {\bibinfo {author} {\bibfnamefont {A.}~\bibnamefont
  {Ribak}}, \bibinfo {author} {\bibfnamefont {R.~M.}\ \bibnamefont {Skiff}},
  \bibinfo {author} {\bibfnamefont {M.}~\bibnamefont {Mograbi}}, \bibinfo
  {author} {\bibfnamefont {P.}~\bibnamefont {Rout}}, \bibinfo {author}
  {\bibfnamefont {M.}~\bibnamefont {Fischer}}, \bibinfo {author} {\bibfnamefont
  {J.}~\bibnamefont {Ruhman}}, \bibinfo {author} {\bibfnamefont
  {K.}~\bibnamefont {Chashka}}, \bibinfo {author} {\bibfnamefont
  {Y.}~\bibnamefont {Dagan}},\ and\ \bibinfo {author} {\bibfnamefont
  {A.}~\bibnamefont {Kanigel}},\ }\bibfield  {title} {\bibinfo {title} {Chiral
  superconductivity in the alternate stacking compound {4Hb-TaS$_2$}},\ }\href
  {https://www.science.org/doi/abs/10.1126/sciadv.aax9480} {\bibfield
  {journal} {\bibinfo  {journal} {Science advances}\ }\textbf {\bibinfo
  {volume} {6}},\ \bibinfo {pages} {eaax9480} (\bibinfo {year}
  {2020})}\BibitemShut {NoStop}%
\bibitem [{\citenamefont {Persky}\ \emph {et~al.}(2022)\citenamefont {Persky},
  \citenamefont {Bj{\o}rlig}, \citenamefont {Feldman}, \citenamefont
  {Almoalem}, \citenamefont {Altman}, \citenamefont {Berg}, \citenamefont
  {Kimchi}, \citenamefont {Ruhman}, \citenamefont {Kanigel},\ and\
  \citenamefont {Kalisky}}]{persky2022}%
  \BibitemOpen
  \bibfield  {author} {\bibinfo {author} {\bibfnamefont {E.}~\bibnamefont
  {Persky}}, \bibinfo {author} {\bibfnamefont {A.~V.}\ \bibnamefont
  {Bj{\o}rlig}}, \bibinfo {author} {\bibfnamefont {I.}~\bibnamefont {Feldman}},
  \bibinfo {author} {\bibfnamefont {A.}~\bibnamefont {Almoalem}}, \bibinfo
  {author} {\bibfnamefont {E.}~\bibnamefont {Altman}}, \bibinfo {author}
  {\bibfnamefont {E.}~\bibnamefont {Berg}}, \bibinfo {author} {\bibfnamefont
  {I.}~\bibnamefont {Kimchi}}, \bibinfo {author} {\bibfnamefont
  {J.}~\bibnamefont {Ruhman}}, \bibinfo {author} {\bibfnamefont
  {A.}~\bibnamefont {Kanigel}},\ and\ \bibinfo {author} {\bibfnamefont
  {B.}~\bibnamefont {Kalisky}},\ }\bibfield  {title} {\bibinfo {title}
  {Magnetic memory and spontaneous vortices in a van der {Waals}
  superconductor},\ }\href {https://doi.org/10.1038/s41586-022-04855-2}
  {\bibfield  {journal} {\bibinfo  {journal} {Nature}\ }\textbf {\bibinfo
  {volume} {607}},\ \bibinfo {pages} {692} (\bibinfo {year}
  {2022})}\BibitemShut {NoStop}%
\bibitem [{\citenamefont {Almoalem}\ \emph {et~al.}(2022)\citenamefont
  {Almoalem}, \citenamefont {Feldman}, \citenamefont {Shlafman}, \citenamefont
  {Yaish}, \citenamefont {Fischer}, \citenamefont {Moshe}, \citenamefont
  {Ruhman},\ and\ \citenamefont {Kanigel}}]{almoalem2022}%
  \BibitemOpen
  \bibfield  {author} {\bibinfo {author} {\bibfnamefont {A.}~\bibnamefont
  {Almoalem}}, \bibinfo {author} {\bibfnamefont {I.}~\bibnamefont {Feldman}},
  \bibinfo {author} {\bibfnamefont {M.}~\bibnamefont {Shlafman}}, \bibinfo
  {author} {\bibfnamefont {Y.~E.}\ \bibnamefont {Yaish}}, \bibinfo {author}
  {\bibfnamefont {M.~H.}\ \bibnamefont {Fischer}}, \bibinfo {author}
  {\bibfnamefont {M.}~\bibnamefont {Moshe}}, \bibinfo {author} {\bibfnamefont
  {J.}~\bibnamefont {Ruhman}},\ and\ \bibinfo {author} {\bibfnamefont
  {A.}~\bibnamefont {Kanigel}},\ }\bibfield  {title} {\bibinfo {title}
  {Evidence of a two-component order parameter in {4Hb-TaS$_2$} in the
  little-parks effect},\ }\href {https://doi.org/10.48550/arXiv.2208.13798}
  {\bibfield  {journal} {\bibinfo  {journal} {arXiv preprint arXiv:2208.13798}\
  } (\bibinfo {year} {2022})}\BibitemShut {NoStop}%
\bibitem [{\citenamefont {Silber}\ \emph {et~al.}(2024)\citenamefont {Silber},
  \citenamefont {Mathimalar}, \citenamefont {Mangel}, \citenamefont {Nayak},
  \citenamefont {Green}, \citenamefont {Avraham}, \citenamefont {Beidenkopf},
  \citenamefont {Feldman}, \citenamefont {Kanigel}, \citenamefont {Klein} \emph
  {et~al.}}]{silber2024}%
  \BibitemOpen
  \bibfield  {author} {\bibinfo {author} {\bibfnamefont {I.}~\bibnamefont
  {Silber}}, \bibinfo {author} {\bibfnamefont {S.}~\bibnamefont {Mathimalar}},
  \bibinfo {author} {\bibfnamefont {I.}~\bibnamefont {Mangel}}, \bibinfo
  {author} {\bibfnamefont {A.}~\bibnamefont {Nayak}}, \bibinfo {author}
  {\bibfnamefont {O.}~\bibnamefont {Green}}, \bibinfo {author} {\bibfnamefont
  {N.}~\bibnamefont {Avraham}}, \bibinfo {author} {\bibfnamefont
  {H.}~\bibnamefont {Beidenkopf}}, \bibinfo {author} {\bibfnamefont
  {I.}~\bibnamefont {Feldman}}, \bibinfo {author} {\bibfnamefont
  {A.}~\bibnamefont {Kanigel}}, \bibinfo {author} {\bibfnamefont
  {A.}~\bibnamefont {Klein}}, \emph {et~al.},\ }\bibfield  {title} {\bibinfo
  {title} {Two-component nematic superconductivity in {4Hb-TaS$_2$}},\ }\href
  {https://doi.org/10.1038/s41467-024-45169-3} {\bibfield  {journal} {\bibinfo
  {journal} {Nature Communications}\ }\textbf {\bibinfo {volume} {15}},\
  \bibinfo {pages} {824} (\bibinfo {year} {2024})}\BibitemShut {NoStop}%
\bibitem [{\citenamefont {Nagata}\ \emph {et~al.}(1992)\citenamefont {Nagata},
  \citenamefont {Aochi}, \citenamefont {Abe}, \citenamefont {Ebisu},
  \citenamefont {Hagino}, \citenamefont {Seki},\ and\ \citenamefont
  {Tsutsumi}}]{nagata1992}%
  \BibitemOpen
  \bibfield  {author} {\bibinfo {author} {\bibfnamefont {S.}~\bibnamefont
  {Nagata}}, \bibinfo {author} {\bibfnamefont {T.}~\bibnamefont {Aochi}},
  \bibinfo {author} {\bibfnamefont {T.}~\bibnamefont {Abe}}, \bibinfo {author}
  {\bibfnamefont {S.}~\bibnamefont {Ebisu}}, \bibinfo {author} {\bibfnamefont
  {T.}~\bibnamefont {Hagino}}, \bibinfo {author} {\bibfnamefont
  {Y.}~\bibnamefont {Seki}},\ and\ \bibinfo {author} {\bibfnamefont
  {K.}~\bibnamefont {Tsutsumi}},\ }\bibfield  {title} {\bibinfo {title}
  {Superconductivity in the layered compound {2H-TaS$_2$}},\ }\href
  {https://doi.org/10.1016/0022-3697(92)90242-6} {\bibfield  {journal}
  {\bibinfo  {journal} {Journal of Physics and Chemistry of Solids}\ }\textbf
  {\bibinfo {volume} {53}},\ \bibinfo {pages} {1259} (\bibinfo {year}
  {1992})}\BibitemShut {NoStop}%
\bibitem [{\citenamefont {Navarro-Moratalla}\ \emph {et~al.}(2016)\citenamefont
  {Navarro-Moratalla}, \citenamefont {Island}, \citenamefont {Manas-Valero},
  \citenamefont {Pinilla-Cienfuegos}, \citenamefont {Castellanos-Gomez},
  \citenamefont {Quereda}, \citenamefont {Rubio-Bollinger}, \citenamefont
  {Chirolli}, \citenamefont {Silva-Guill{\'e}n}, \citenamefont {Agra{\"\i}t}
  \emph {et~al.}}]{navarro2016enhanced}%
  \BibitemOpen
  \bibfield  {author} {\bibinfo {author} {\bibfnamefont {E.}~\bibnamefont
  {Navarro-Moratalla}}, \bibinfo {author} {\bibfnamefont {J.~O.}\ \bibnamefont
  {Island}}, \bibinfo {author} {\bibfnamefont {S.}~\bibnamefont
  {Manas-Valero}}, \bibinfo {author} {\bibfnamefont {E.}~\bibnamefont
  {Pinilla-Cienfuegos}}, \bibinfo {author} {\bibfnamefont {A.}~\bibnamefont
  {Castellanos-Gomez}}, \bibinfo {author} {\bibfnamefont {J.}~\bibnamefont
  {Quereda}}, \bibinfo {author} {\bibfnamefont {G.}~\bibnamefont
  {Rubio-Bollinger}}, \bibinfo {author} {\bibfnamefont {L.}~\bibnamefont
  {Chirolli}}, \bibinfo {author} {\bibfnamefont {J.~A.}\ \bibnamefont
  {Silva-Guill{\'e}n}}, \bibinfo {author} {\bibfnamefont {N.}~\bibnamefont
  {Agra{\"\i}t}}, \emph {et~al.},\ }\bibfield  {title} {\bibinfo {title}
  {Enhanced superconductivity in atomically thin {TaS$_2$}},\ }\href
  {https://doi.org/10.1038/ncomms11043} {\bibfield  {journal} {\bibinfo
  {journal} {Nature communications}\ }\textbf {\bibinfo {volume} {7}},\
  \bibinfo {pages} {11043} (\bibinfo {year} {2016})}\BibitemShut {NoStop}%
\bibitem [{\citenamefont {Law}\ and\ \citenamefont {Lee}(2017)}]{law2017}%
  \BibitemOpen
  \bibfield  {author} {\bibinfo {author} {\bibfnamefont {K.~T.}\ \bibnamefont
  {Law}}\ and\ \bibinfo {author} {\bibfnamefont {P.~A.}\ \bibnamefont {Lee}},\
  }\bibfield  {title} {\bibinfo {title} {{1T-TaS$_2$} as a quantum spin
  liquid},\ }\href {https://doi.org/10.1073/pnas.1706769114} {\bibfield
  {journal} {\bibinfo  {journal} {Proceedings of the National Academy of
  Sciences}\ }\textbf {\bibinfo {volume} {114}},\ \bibinfo {pages} {6996}
  (\bibinfo {year} {2017})}\BibitemShut {NoStop}%
\bibitem [{\citenamefont {He}\ \emph {et~al.}(2018)\citenamefont {He},
  \citenamefont {Xu}, \citenamefont {Chen}, \citenamefont {Law},\ and\
  \citenamefont {Lee}}]{he2018}%
  \BibitemOpen
  \bibfield  {author} {\bibinfo {author} {\bibfnamefont {W.-Y.}\ \bibnamefont
  {He}}, \bibinfo {author} {\bibfnamefont {X.~Y.}\ \bibnamefont {Xu}}, \bibinfo
  {author} {\bibfnamefont {G.}~\bibnamefont {Chen}}, \bibinfo {author}
  {\bibfnamefont {K.~T.}\ \bibnamefont {Law}},\ and\ \bibinfo {author}
  {\bibfnamefont {P.~A.}\ \bibnamefont {Lee}},\ }\bibfield  {title} {\bibinfo
  {title} {Spinon fermi surface in a cluster mott insulator model on a
  triangular lattice and possible application to
  {$1\mathit{T}{-}\text{TaS}_2$}},\ }\href
  {https://doi.org/10.1103/PhysRevLett.121.046401} {\bibfield  {journal}
  {\bibinfo  {journal} {Physical review letters}\ }\textbf {\bibinfo {volume}
  {121}},\ \bibinfo {pages} {046401} (\bibinfo {year} {2018})}\BibitemShut
  {NoStop}%
\bibitem [{\citenamefont {Ruan}\ \emph {et~al.}(2021)\citenamefont {Ruan},
  \citenamefont {Chen}, \citenamefont {Tang}, \citenamefont {Hwang},
  \citenamefont {Tsai}, \citenamefont {Lee}, \citenamefont {Wu}, \citenamefont
  {Ryu}, \citenamefont {Kahn}, \citenamefont {Liou} \emph
  {et~al.}}]{ruan2021evidence}%
  \BibitemOpen
  \bibfield  {author} {\bibinfo {author} {\bibfnamefont {W.}~\bibnamefont
  {Ruan}}, \bibinfo {author} {\bibfnamefont {Y.}~\bibnamefont {Chen}}, \bibinfo
  {author} {\bibfnamefont {S.}~\bibnamefont {Tang}}, \bibinfo {author}
  {\bibfnamefont {J.}~\bibnamefont {Hwang}}, \bibinfo {author} {\bibfnamefont
  {H.-Z.}\ \bibnamefont {Tsai}}, \bibinfo {author} {\bibfnamefont {R.~L.}\
  \bibnamefont {Lee}}, \bibinfo {author} {\bibfnamefont {M.}~\bibnamefont
  {Wu}}, \bibinfo {author} {\bibfnamefont {H.}~\bibnamefont {Ryu}}, \bibinfo
  {author} {\bibfnamefont {S.}~\bibnamefont {Kahn}}, \bibinfo {author}
  {\bibfnamefont {F.}~\bibnamefont {Liou}}, \emph {et~al.},\ }\bibfield
  {title} {\bibinfo {title} {Evidence for quantum spin liquid behaviour in
  single-layer {1T-TaSe$_2$} from scanning tunnelling microscopy},\ }\href
  {https://doi.org/10.1038/s41567-021-01321-0} {\bibfield  {journal} {\bibinfo
  {journal} {Nature Physics}\ }\textbf {\bibinfo {volume} {17}},\ \bibinfo
  {pages} {1154} (\bibinfo {year} {2021})}\BibitemShut {NoStop}%
\bibitem [{\citenamefont {Zhang}\ \emph {et~al.}(2024)\citenamefont {Zhang},
  \citenamefont {He}, \citenamefont {Zhang}, \citenamefont {Chen},
  \citenamefont {Jia}, \citenamefont {Hou}, \citenamefont {Ji}, \citenamefont
  {Yang}, \citenamefont {Zhang}, \citenamefont {Liu} \emph
  {et~al.}}]{houquantum2024}%
  \BibitemOpen
  \bibfield  {author} {\bibinfo {author} {\bibfnamefont {Q.}~\bibnamefont
  {Zhang}}, \bibinfo {author} {\bibfnamefont {W.-Y.}\ \bibnamefont {He}},
  \bibinfo {author} {\bibfnamefont {Y.}~\bibnamefont {Zhang}}, \bibinfo
  {author} {\bibfnamefont {Y.}~\bibnamefont {Chen}}, \bibinfo {author}
  {\bibfnamefont {L.}~\bibnamefont {Jia}}, \bibinfo {author} {\bibfnamefont
  {Y.}~\bibnamefont {Hou}}, \bibinfo {author} {\bibfnamefont {H.}~\bibnamefont
  {Ji}}, \bibinfo {author} {\bibfnamefont {H.}~\bibnamefont {Yang}}, \bibinfo
  {author} {\bibfnamefont {T.}~\bibnamefont {Zhang}}, \bibinfo {author}
  {\bibfnamefont {L.}~\bibnamefont {Liu}}, \emph {et~al.},\ }\bibfield  {title}
  {\bibinfo {title} {Quantum spin liquid signatures in monolayer
  {1T-NbSe$_2$}},\ }\href {https://doi.org/10.1038/s41467-024-46612-1}
  {\bibfield  {journal} {\bibinfo  {journal} {Nature Communications}\ }\textbf
  {\bibinfo {volume} {15}},\ \bibinfo {pages} {2336} (\bibinfo {year}
  {2024})}\BibitemShut {NoStop}%
\bibitem [{\citenamefont {Hu}\ \emph {et~al.}(2016)\citenamefont {Hu},
  \citenamefont {Gong},\ and\ \citenamefont {Sheng}}]{hu2016}%
  \BibitemOpen
  \bibfield  {author} {\bibinfo {author} {\bibfnamefont {W.-J.}\ \bibnamefont
  {Hu}}, \bibinfo {author} {\bibfnamefont {S.-S.}\ \bibnamefont {Gong}},\ and\
  \bibinfo {author} {\bibfnamefont {D.}~\bibnamefont {Sheng}},\ }\bibfield
  {title} {\bibinfo {title} {Variational monte carlo study of chiral spin
  liquid in quantum antiferromagnet on the triangular lattice},\ }\href
  {https://doi.org/10.1103/PhysRevB.94.075131} {\bibfield  {journal} {\bibinfo
  {journal} {Physical Review B}\ }\textbf {\bibinfo {volume} {94}},\ \bibinfo
  {pages} {075131} (\bibinfo {year} {2016})}\BibitemShut {NoStop}%
\bibitem [{\citenamefont {Szasz}\ \emph {et~al.}(2020)\citenamefont {Szasz},
  \citenamefont {Motruk}, \citenamefont {Zaletel},\ and\ \citenamefont
  {Moore}}]{szasz2020}%
  \BibitemOpen
  \bibfield  {author} {\bibinfo {author} {\bibfnamefont {A.}~\bibnamefont
  {Szasz}}, \bibinfo {author} {\bibfnamefont {J.}~\bibnamefont {Motruk}},
  \bibinfo {author} {\bibfnamefont {M.~P.}\ \bibnamefont {Zaletel}},\ and\
  \bibinfo {author} {\bibfnamefont {J.~E.}\ \bibnamefont {Moore}},\ }\bibfield
  {title} {\bibinfo {title} {Chiral spin liquid phase of the triangular lattice
  hubbard model: A density matrix renormalization group study},\ }\href
  {https://doi.org/10.1103/PhysRevX.10.021042} {\bibfield  {journal} {\bibinfo
  {journal} {Physical Review X}\ }\textbf {\bibinfo {volume} {10}},\ \bibinfo
  {pages} {021042} (\bibinfo {year} {2020})}\BibitemShut {NoStop}%
\bibitem [{\citenamefont {Crippa}\ \emph {et~al.}(2024)\citenamefont {Crippa},
  \citenamefont {Bae}, \citenamefont {Wunderlich}, \citenamefont {Mazin},
  \citenamefont {Yan}, \citenamefont {Sangiovanni}, \citenamefont {Wehling},\
  and\ \citenamefont {Valent{\'\i}}}]{crippa2024}%
  \BibitemOpen
  \bibfield  {author} {\bibinfo {author} {\bibfnamefont {L.}~\bibnamefont
  {Crippa}}, \bibinfo {author} {\bibfnamefont {H.}~\bibnamefont {Bae}},
  \bibinfo {author} {\bibfnamefont {P.}~\bibnamefont {Wunderlich}}, \bibinfo
  {author} {\bibfnamefont {I.~I.}\ \bibnamefont {Mazin}}, \bibinfo {author}
  {\bibfnamefont {B.}~\bibnamefont {Yan}}, \bibinfo {author} {\bibfnamefont
  {G.}~\bibnamefont {Sangiovanni}}, \bibinfo {author} {\bibfnamefont
  {T.}~\bibnamefont {Wehling}},\ and\ \bibinfo {author} {\bibfnamefont
  {R.}~\bibnamefont {Valent{\'\i}}},\ }\bibfield  {title} {\bibinfo {title}
  {Heavy fermions vs doped mott physics in heterogeneous {Ta}-dichalcogenide
  bilayers},\ }\href {https://doi.org/10.1038/s41467-024-45392-y} {\bibfield
  {journal} {\bibinfo  {journal} {Nature Communications}\ }\textbf {\bibinfo
  {volume} {15}},\ \bibinfo {pages} {1357} (\bibinfo {year}
  {2024})}\BibitemShut {NoStop}%
\bibitem [{\citenamefont {Wen}\ \emph {et~al.}(2021)\citenamefont {Wen},
  \citenamefont {Gao}, \citenamefont {Xie}, \citenamefont {Zhang},
  \citenamefont {Kong}, \citenamefont {Wang}, \citenamefont {Jiang},
  \citenamefont {Luo}, \citenamefont {Li}, \citenamefont {Lu}, \citenamefont
  {Sun},\ and\ \citenamefont {Yan}}]{wen2021}%
  \BibitemOpen
  \bibfield  {author} {\bibinfo {author} {\bibfnamefont {C.}~\bibnamefont
  {Wen}}, \bibinfo {author} {\bibfnamefont {J.}~\bibnamefont {Gao}}, \bibinfo
  {author} {\bibfnamefont {Y.}~\bibnamefont {Xie}}, \bibinfo {author}
  {\bibfnamefont {Q.}~\bibnamefont {Zhang}}, \bibinfo {author} {\bibfnamefont
  {P.}~\bibnamefont {Kong}}, \bibinfo {author} {\bibfnamefont {J.}~\bibnamefont
  {Wang}}, \bibinfo {author} {\bibfnamefont {Y.}~\bibnamefont {Jiang}},
  \bibinfo {author} {\bibfnamefont {X.}~\bibnamefont {Luo}}, \bibinfo {author}
  {\bibfnamefont {J.}~\bibnamefont {Li}}, \bibinfo {author} {\bibfnamefont
  {W.}~\bibnamefont {Lu}}, \bibinfo {author} {\bibfnamefont {Y.-P.}\
  \bibnamefont {Sun}},\ and\ \bibinfo {author} {\bibfnamefont {S.}~\bibnamefont
  {Yan}},\ }\bibfield  {title} {\bibinfo {title} {Roles of the narrow
  electronic band near the fermi level in
  {$1\mathit{T}{-}\text{TaS}_2$}-related layered materials},\ }\href
  {https://doi.org/10.1103/PhysRevLett.126.256402} {\bibfield  {journal}
  {\bibinfo  {journal} {Phys. Rev. Lett.}\ }\textbf {\bibinfo {volume} {126}},\
  \bibinfo {pages} {256402} (\bibinfo {year} {2021})}\BibitemShut {NoStop}%
\bibitem [{\citenamefont {Kumar~Nayak}\ \emph {et~al.}(2023)\citenamefont
  {Kumar~Nayak}, \citenamefont {Steinbok}, \citenamefont {Roet}, \citenamefont
  {Koo}, \citenamefont {Feldman}, \citenamefont {Almoalem}, \citenamefont
  {Kanigel}, \citenamefont {Yan}, \citenamefont {Rosch}, \citenamefont
  {Avraham} \emph {et~al.}}]{nayak2023}%
  \BibitemOpen
  \bibfield  {author} {\bibinfo {author} {\bibfnamefont {A.}~\bibnamefont
  {Kumar~Nayak}}, \bibinfo {author} {\bibfnamefont {A.}~\bibnamefont
  {Steinbok}}, \bibinfo {author} {\bibfnamefont {Y.}~\bibnamefont {Roet}},
  \bibinfo {author} {\bibfnamefont {J.}~\bibnamefont {Koo}}, \bibinfo {author}
  {\bibfnamefont {I.}~\bibnamefont {Feldman}}, \bibinfo {author} {\bibfnamefont
  {A.}~\bibnamefont {Almoalem}}, \bibinfo {author} {\bibfnamefont
  {A.}~\bibnamefont {Kanigel}}, \bibinfo {author} {\bibfnamefont
  {B.}~\bibnamefont {Yan}}, \bibinfo {author} {\bibfnamefont {A.}~\bibnamefont
  {Rosch}}, \bibinfo {author} {\bibfnamefont {N.}~\bibnamefont {Avraham}},
  \emph {et~al.},\ }\bibfield  {title} {\bibinfo {title} {First-order quantum
  phase transition in the hybrid metal--mott insulator transition metal
  dichalcogenide {4Hb-TaS$_2$}},\ }\href
  {https://doi.org/10.1073/pnas.2304274120} {\bibfield  {journal} {\bibinfo
  {journal} {Proceedings of the National Academy of Sciences}\ }\textbf
  {\bibinfo {volume} {120}},\ \bibinfo {pages} {e2304274120} (\bibinfo {year}
  {2023})}\BibitemShut {NoStop}%
\bibitem [{\citenamefont {Fischer}\ \emph {et~al.}(2023)\citenamefont
  {Fischer}, \citenamefont {Lee},\ and\ \citenamefont {Ruhman}}]{fischer2023}%
  \BibitemOpen
  \bibfield  {author} {\bibinfo {author} {\bibfnamefont {M.~H.}\ \bibnamefont
  {Fischer}}, \bibinfo {author} {\bibfnamefont {P.~A.}\ \bibnamefont {Lee}},\
  and\ \bibinfo {author} {\bibfnamefont {J.}~\bibnamefont {Ruhman}},\
  }\bibfield  {title} {\bibinfo {title} {Mechanism for $\ensuremath{\pi}$ phase
  shifts in little-parks experiments: Application to
  {$4\mathit{Hb}{-}\text{TaS}_2$} and to {$2\mathit{H}{-}\text{TaS}_{2}$}
  intercalated with chiral molecules},\ }\href
  {https://doi.org/10.1103/PhysRevB.108.L180505} {\bibfield  {journal}
  {\bibinfo  {journal} {Phys. Rev. B}\ }\textbf {\bibinfo {volume} {108}},\
  \bibinfo {pages} {L180505} (\bibinfo {year} {2023})}\BibitemShut {NoStop}%
\bibitem [{\citenamefont {Hammer}\ \emph {et~al.}(2007)\citenamefont {Hammer},
  \citenamefont {Cuevas}, \citenamefont {Bergeret},\ and\ \citenamefont
  {Belzig}}]{Hammer:prb07}%
  \BibitemOpen
  \bibfield  {author} {\bibinfo {author} {\bibfnamefont {J.~C.}\ \bibnamefont
  {Hammer}}, \bibinfo {author} {\bibfnamefont {J.~C.}\ \bibnamefont {Cuevas}},
  \bibinfo {author} {\bibfnamefont {F.~S.}\ \bibnamefont {Bergeret}},\ and\
  \bibinfo {author} {\bibfnamefont {W.}~\bibnamefont {Belzig}},\ }\bibfield
  {title} {\bibinfo {title} {Density of states and supercurrent in diffusive
  sns junctions: Roles of nonideal interfaces and spin-flip scattering},\
  }\href {https://doi.org/10.1103/PhysRevB.76.064514} {\bibfield  {journal}
  {\bibinfo  {journal} {Phys. Rev. B}\ }\textbf {\bibinfo {volume} {76}},\
  \bibinfo {pages} {064514} (\bibinfo {year} {2007})}\BibitemShut {NoStop}%
\bibitem [{\citenamefont {Cherkez}\ \emph {et~al.}(2014)\citenamefont
  {Cherkez}, \citenamefont {Cuevas}, \citenamefont {Brun}, \citenamefont
  {Cren}, \citenamefont {M\'enard}, \citenamefont {Debontridder}, \citenamefont
  {Stolyarov},\ and\ \citenamefont {Roditchev}}]{Cherkez:prx14}%
  \BibitemOpen
  \bibfield  {author} {\bibinfo {author} {\bibfnamefont {V.}~\bibnamefont
  {Cherkez}}, \bibinfo {author} {\bibfnamefont {J.~C.}\ \bibnamefont {Cuevas}},
  \bibinfo {author} {\bibfnamefont {C.}~\bibnamefont {Brun}}, \bibinfo {author}
  {\bibfnamefont {T.}~\bibnamefont {Cren}}, \bibinfo {author} {\bibfnamefont
  {G.}~\bibnamefont {M\'enard}}, \bibinfo {author} {\bibfnamefont
  {F.}~\bibnamefont {Debontridder}}, \bibinfo {author} {\bibfnamefont {V.~S.}\
  \bibnamefont {Stolyarov}},\ and\ \bibinfo {author} {\bibfnamefont
  {D.}~\bibnamefont {Roditchev}},\ }\bibfield  {title} {\bibinfo {title}
  {Proximity effect between two superconductors spatially resolved by scanning
  tunneling spectroscopy},\ }\href {https://doi.org/10.1103/PhysRevX.4.011033}
  {\bibfield  {journal} {\bibinfo  {journal} {Phys. Rev. X}\ }\textbf {\bibinfo
  {volume} {4}},\ \bibinfo {pages} {011033} (\bibinfo {year}
  {2014})}\BibitemShut {NoStop}%
\bibitem [{\citenamefont {Kuprianov}\ and\ \citenamefont
  {Lukichev}(1988)}]{kuprianov1988}%
  \BibitemOpen
  \bibfield  {author} {\bibinfo {author} {\bibfnamefont {M.~Y.}\ \bibnamefont
  {Kuprianov}}\ and\ \bibinfo {author} {\bibfnamefont {V.}~\bibnamefont
  {Lukichev}},\ }\bibfield  {title} {\bibinfo {title} {Influence of boundary
  transparency on the critical current of “dirty” ss’s structures},\
  }\href@noop {} {\bibfield  {journal} {\bibinfo  {journal} {Zh. Eksp. Teor.
  Fiz}\ }\textbf {\bibinfo {volume} {94}},\ \bibinfo {pages} {139} (\bibinfo
  {year} {1988})}\BibitemShut {NoStop}%
\end{thebibliography}%
\end{document}